\documentclass[superscriptaddress,a4paper,tightenlines,nofootinbib,floatfix]{article}
\pdfoutput=1

\usepackage{amsmath,amssymb,jheppub,mathtools}
\usepackage{graphicx}
\usepackage{dcolumn}
\usepackage{bm}
\usepackage{enumitem}
\usepackage[mathlines]{lineno}
\usepackage{xcolor}
\usepackage{multirow}
\usepackage{url}
\usepackage{float,slashed}
\usepackage{soul}
\usepackage{makecell}
\usepackage{subcaption}

\usepackage[normalem]{ulem}

\renewcommand{\texttt}[1]{%
  \begingroup
  \ttfamily
  \begingroup\lccode`~=`/\lowercase{\endgroup\def~}{/\discretionary{}{}{}}%
  \begingroup\lccode`~=`[\lowercase{\endgroup\def~}{[\discretionary{}{}{}}%
  \begingroup\lccode`~=`.\lowercase{\endgroup\def~}{.\discretionary{}{}{}}%
  \catcode`/=\active\catcode`[=\active\catcode`.=\active
  \scantokens{#1\noexpand}%
  \endgroup
}
\def\issue(#1,#2,#3){{\bf #1}, #2 (#3)}

\catcode`\@=11
\def\lsim{\mathrel{\mathpalette\@versim<}}
\def\gsim{\mathrel{\mathpalette\@versim>}}
\def\@versim#1#2{\vcenter{\offinterlineskip
\ialign{$\m@th#1\hfil##\hfil$\crcr#2\crcr\sim\crcr } }}
\catcode`\@=12

\catcode`@=12

\newcommand{\met}{$\cancel E_T$}
\newcommand{\newc}{\newcommand}
\newc{\wt}{\widetilde}
\newc{\ra}{\rightarrow}
\def\beq {\begin{equation}}
\def\eeq {\end{equation}}
\def\bi {\begin{itemize}}
\def\ei {\end{itemize}}
\def\bea {\begin{eqnarray}}
\def\eea {\end{eqnarray}}

\def \met{\rm E{\!\!\!/}_T} 

\def \PMETV{\overrightarrow{\rm p}{\!\!\!\!/}_T}
\newcommand{\br}{\begin{eqnarray}}
\newcommand{\er}{\end{eqnarray}}
\newcommand{\be}{\begin{equation}}
\newcommand{\ee}{\end{equation}}

\newcommand{\ch}{\widetilde \chi^\pm}

\def \ch2p {{\wt\chi_2^+}}

\def \ch2m {{\wt\chi_2^-}}

\def \chonepm{{\wt\chi_1}^{\pm}}
\def \chonemp{{\wt\chi_1}^{\mp}}
\def \mchonepm{m_{\chonepm}}

\newc{\dmchi}{\Delta m_{\wt\chi}}

\def \chtwopm{{\wt\chi_2}^{\pm}}

\def \lspi{\wt\chi_i^0}
\def \mlspi{m_{\lspi}}

\def \lspone{\wt\chi_1^0}

\def \lsptwo{\wt\chi_2^0}

\def \lspthree{\wt\chi_3^0}

\def \lspfour{\wt\chi_4^0}

\def \lspfive{\wt\chi_5^0}

\title{Exploring the Singlino-dominated Thermal Neutralino Dark Matter in the $Z_3$ invariant NMSSM}
\author[1]{Amandip De}
\author[2]{Amit Adhikary}
\author[3]{Rahool Kumar Barman}
\author[4]{Biplob Bhattacherjee}
\author[4]{Rohini M. Godbole}

\affiliation[1]{Department of Physics, SRM University-AP, Amaravati 522240, Andhra Pradesh, India}
\affiliation[2]{Aix Marseille Univ., Universit\'e de Toulon, CNRS, CPT, IPhU, Marseille, France}
\affiliation[3]{Kavli IPMU (WPI), UTIAS, The University of Tokyo, Kashiwa, Chiba 277-8583, Japan}
\affiliation[4]{Centre for High Energy Physics, Indian Institute of Science, Bangalore 560012, India}

\emailAdd{amit.adhikary@cpt.univ-mrs.fr}
\emailAdd{rahool.barman@ipmu.jp}
\emailAdd{biplob@iisc.ac.in}
\emailAdd{rohini@iisc.ac.in} 
\emailAdd{deamandip93@gmail.com}

\abstract{We examine the parameter space of the Next to Minimal Supersymmetric Standard Model~(NMSSM) with Singlino-dominated neutralino $\lspone$ as the lightest supersymmetric particle~(LSP). Our study focuses on identifying the regions within this parameter space that produce a thermal relic abundance of $\lspone$ smaller than the observed cold dark matter relic density while remaining consistent with constraints from  LEP measurements, low-energy experiments, Higgs measurements, LHC data, and dark matter direct detection experiments. We identify the dominant annihilation modes of the LSP neutralino across varying LSP mass ranges $\sim \mathcal{O}(1)-\mathcal{O}(10^{3})~$GeV. Furthermore, we conduct a benchmark study to assess the production rates of triple-boson final states emerging from direct electroweakino pair production at the LHC. Drawing insights from these findings, we perform a detailed collider analysis to explore the future potential of probing the triple-boson final states involving a light Higgs boson at the high-luminosity LHC~(HL-LHC).}
\date{\today}

\begin{document}
\maketitle
\section{Introduction}
\label{sec:intro}

The nature of Dark Matter (DM) remains a central mystery in contemporary particle physics, particularly within the realm of Beyond Standard Model (BSM) physics. Unveiling the properties of DM candidates is a key objective, driving extensive experimental efforts that employ direct, indirect \cite{Hooper:2008sn, Feng:2010gw} and collider searches \cite{Goodman:2010yf, Fox:2011pm}. Among the proposed DM candidates, a Weakly Interacting Massive Particle (WIMP), an example of non-baryonic Cold Dark Matter (CDM) candidate interacting via electroweak coupling with a mass around the electroweak scale, can naturally predict a relic density that aligns very closely to the measured value of $\Omega h^2 = 0.120 \pm 0.001$ in the PLACK experiment at 68\% C.L. \cite{Planck:2018vyg}. WIMPs are one of the favourable DM candidates since they can be produced at the thermal equilibrium conditions prevailing in this surprising agreement, referred to as the `WIMP miracle' \cite{Arcadi:2017kky, Dutra:2019gqz, Drees:2007kk}. The non-relativistic nature and weak interactions of WIMPs lead to two primary categories for DM-nucleon scattering: spin-independent (SI) and spin-dependent (SD). The recent LUX-ZEPLIN (LZ) experiment has reported the most stringent limit of DM-nucleon SI elastic scattering cross-section ($\sigma_{SI}$) at $9.2 \times 10^{-48}~\mathrm{cm^2}$ at the $90\%$ C.L. corresponding to DM particle mass at 36 GeV \cite{PhysRevLett.131.041002}. While the SD WIMP-nuclear cross-section has the minimum limit for 30 and 25 GeV WIMPs at $1.49\times 10^{-42}$ and $3.2\times 10^{-41} \mathrm{cm^2}$ from LZ and PICO-60 $C_3 F_8$ experiments for SD WIMP-neutron ($\sigma_{SD_n}$) and SD WIMP-proton ($\sigma_{SD_p}$) cross-sections \cite{PhysRevLett.131.041002, PhysRevD.100.022001}. LZ for 1000 live day run has projected an increase by around an order of magnitude in these cross-sections \cite{LZ:2018qzl}. These strong limits on SI and SD DM-nucleon scattering indicate the feeble nature of the DM and nucleon interactions, which poses significant experimental challenges \cite{Cao:2019qng}. This necessitates exploring alternative avenues for DM detection, such as searching for missing energy signatures in experiments at the LHC.   

The absence of any DM candidate in the SM strongly suggests looking for possible BSM scenarios to explain the observed dark matter. A very well-motivated model, Supersymmetry, can provide a possible dark matter candidate, the lightest neutralino, within R-parity conserved scenario  \cite{doi:10.1142/4001, baer_tata_2006}. The neutralino sector in Minimal Supersymmetric Standard Model (MSSM) involves the mixing of Bino ($\tilde{B}$), Wino ($\tilde{W}$), and neutral Higgsinos ($\tilde{H}_u$ and $\tilde{H}_d$), to produce four neutralinos. The lightest neutralino ($\lspone$), denoted as the lightest supersymmetric particle (LSP), a Majorana spin 1/2 particle, is usually the most popular candidate for DM. In phenomenological MSSM (pMSSM), LSP mass must exceed 34 GeV to avoid over-abundant relic density ~\cite{Calibbi:2013poa, Belanger:2013pna, Barman:2017swy, Cahill-Rowley:2014twa, Cao:2015efs}. For single-component thermal DM in the general MSSM, Higgsinos are favoured to be massive ($\gtrsim$ 1 TeV) to achieve the correct DM abundance ~\cite{Arkani-Hamed:2006wnf, Baer:2016ucr, Chakraborti:2017dpu}. This avoids an overly large annihilation rate unless stops are very heavy \cite{Huang:2017kdh, Badziak:2017the} or the Higgsino mass parameter ($\mu$) is negative \cite{Abdughani:2017dqs}. In the low-mass limit ($\lesssim$ 1 TeV), a Bino-dominated LSP with significant co-annihilation from Wino-like neutralinos is preferred to match the observed relic density. The parameter $\mu$ receives constraints from both the theory and experimental front. Moreover, a large value of Higgsino mass parameter $\mu$ far above the value of $M_{Z}$ is disfavored in the MSSM since it is not {\it natural} and can introduce a ``little fine-tune problem''~\cite{Casas:2003jx, vanBeekveld:2019tqp, Ellwanger:2011mu}. This can be evaded in a model featuring a singlet extension of the MSSM, the Next-to-Minimal Supersymmetric Standard Model (NMSSM) \cite{Ellwanger:1993xa, Ellwanger:2009dp, Maniatis:2009re}, which introduces a singlet Higgs field in addition to two Higgs doublets of the MSSM. In NMSSM, the $\mu$ parameter is generated dynamically when the singlet field develops a vacuum expectation value~($vev$) $v_S$, with $Z_3$ symmetry in the NMSSM, $\mu = \lambda v_S$ at the electroweak scale. 

The NMSSM introduces an extended Higgs sector with seven Higgs bosons: three CP-even, two CP-odd neutrals, and two charged ones. Interestingly, this allows for a light singlet-dominated scalar ($H_1$) or singlet-dominated pseudoscalar Higgs ($A_1$) potentially lighter than the SM-like Higgs ($H_{SM}$) satisfying the collider constraints \cite{Huang:2013ima, Ellwanger:2014hia, Baum:2017enm, Abdallah:2019znp, Guchait:2020wqn, Ellwanger:2024vvs}. Likewise, the Singlino, fermionic superpartner of singlet field, expands the neutralino sector to five states, where the lightest of these neutralinos remains a viable candidate for DM \cite{Cao:2021ljw, Zhou:2021pit, Baum:2019uzg, Domingo:2018ykx, Cao:2018rix, Cao:2022ovk, Adhikary:2022pni, Cao:2022htd, Meng:2024lmi}. The singlet-like light scalars ($H_1/A_1$) are important to our current study, where along with it, a Singlino-dominated LSP containing critical amount of Higgsino admixture for moderate values of $\mu$ \cite{Xiang:2016ndq, Ellwanger:2016sur, Ellwanger:2018zxt, Kim:2014noa} can provide under-abundant relic. In exceedingly low mass regions, this scenario offers new annihilation funnels via light Higgs along with correct Singlino admixture in LSP to satisfy the limits of DM DD, which was not possible in MSSM \cite{Ellwanger:2016sur, Ellwanger:2018zxt}. In the relatively high mass regions, the usual co-annihilation mechanism dominates where Higgsino admixture allows the LSP to efficiently annihilate in the early universe and, in turn, increase the interaction strength of LSP, making it more detectable in DM DD experiments. Moreover, considering a slight Bino content in the LSP for suitably small $M_1$, could provide us with a viable Singlino-dominated DM candidate, which allows funnels featuring SM Higgs boson, Z-boson and even the light singlet-like Higgs \cite{Das:2012rr, Abdallah:2019znp}. Given a low-mass Singlino-like DM candidate, we primarily aim to study if this type of scenario is still compatible with the DM DD and collider searches. Then, look for possible interesting channels involving decays to Singlino-like LSP and their sensitivity of detection at high luminosity LHC in the framework of NMSSM. 

Usually, the interesting channels involve cascade decays of MSSM-like NLSPs to Singlino-like LSP where the decays of heavier neutralinos and charginos proceed via the emission of on/off shell gauge and Higgs bosons $Z/W^\pm/H_{SM}/H_1/A_1$ when the sparticles in the model are heavy, and the electroweakinos do not form a compressed spectrum. Particularly, the condition $m_{LSP} < M_1 < \mu$ in NMSSM facilitate typical decays $\chonepm \to \lspone/\lsptwo~W^\pm, \tilde{\chi}_i^0 \to Z/H_{SM}/H_1/A_1~\lspone$. In this work, we only include the triple-boson channels \cite{ATLAS:2019dny, ATLAS:2022xnu, CMS:2020hjs, CMS-PAS-SMP-19-014} from the Higgsino-like chargino-neutralino production, which is otherwise not native to MSSM. These triple-bosons are usually detected via their leptonic decays, leading to a signal final state consisting of $\met$ and leptons. Some specific consequences of such possibilities incorporate light singlet Higgs producing collimated decays to $b\bar{b}/\tau\tau/W^\pm W^\pm/\gamma\gamma$. We study one such possibility where $H_1$ decays to $b\bar{b}$ and appears as a single fatjet. A few studies have been performed at the LHC to search light scalar in decays $\tau\tau$ \cite{Cerdeno:2013qta}, $b\bar{b}$ \cite{Dutta:2014hma, Guchait:2020wqn}, $\gamma\gamma$ \cite{Domingo:2016yih, Ellwanger:2016wfe, Domingo:2018ykx}. To summarize, we look for a signal final state consisting of a Higgs fatjet, leptons plus $\met$ and predict the sensitivity at HL-LHC. 

The paper is organized as follows. In Sec.~\ref{sec:param_space}, we revisit the Higgs and electroweakino sectors in the NMSSM framework. We discuss the range of parameters for our scan,  relevant collider and astrophysical constraints, and their impact on the parameter space of our interest in Sec.~\ref{sec:constraints}. We examine the different DM annihilation modes responsible for effective DM dilution in the early universe, through which consistency with relic density constraints is achieved in Sec.~\ref{sec:annLSP}. In Sec.~\ref{sec:future_prospects}, we discuss the benchmark scenario adopted for analyzing the HL-LHC potential for direct electroweakino pair production in the triple boson $+\met$ final state. The details of the collider search are presented in Sec.~\ref{sec:collider}. We conclude in Sec.~\ref{sec:conclusion}.    

\section{The NMSSM framework}
\label{sec:param_space}

We consider the ${\rm Z_3}$-invariant NMSSM with the superpotential~\cite{Ellwanger:2009dp},
\begin{equation}
W_{\rm NMSSM} =  W_{\rm MSSM}|_{\mu=0} + \lambda \widehat{S}\widehat{H}_u.\widehat{H}_d + \frac{\kappa}{3}\widehat{S}^3,
\label{eq:superpot}
\end{equation}
where, $W_{\rm MSSM}|_{\mu=0}$ denotes the MSSM superpotential without the $\mu-$term, $\widehat{H}_u$ and $\widehat{H}_d$ are the two doublet Higgs superfields similar to that in MSSM, and $\widehat{S}$ is a gauge singlet Higgs superfield. The dimensionless parameters, ``$\lambda$'' and ``$\kappa$'', control the mixing between the doublet and singlet Higgs superfields and self-coupling of the singlet superfield, respectively. The superpotential in Eq.~\eqref{eq:superpot} provides a solution to the MSSM $\mu$-problem by generating an effective $\mu$-term, $\mu = \lambda v_{S}$, when, $\widehat{S}$ develops a vacuum expectation value $v_S$. 

The Higgs scalar potential $V_{Higgs}$ receives contributions from the soft SUSY breaking terms $V_{soft}$~\cite{Ellwanger:2009dp},
\begin{equation}
\begin{split}
   V_{soft} = & ~m^2_{H_u}|H_u|^2 + m^2_{H_d}|H_d|^2 + m^2_{S}|S|^2 \\ 
   &+ \left( \lambda A_\lambda S H_u\cdot H_d  + \frac{1}{3} \kappa A_k S^3 + \text{h.c.} \right),
\end{split}
   \label{Eqn:v_Soft_term}
\end{equation}
where, $m_{H_u}$, $m_{H_d}$, $m_{S}$ are the soft breaking Higgs masses and $A_{\lambda}$, $A_{\kappa}$ are the trilinear couplings with dimensions of mass. The full scalar Higgs potential is obtained by combining $V_{soft}$ with the $F$ and $D$ terms, $V_{Higgs} = V_{soft} + V_D + V_F$, with,
\begin{equation}
\begin{split}
    &V_F = |\lambda H_u\cdot H_d + \kappa S^2|^2 + \lambda^2 |S|^2 \left( H^\dagger_u H_u  + H^\dagger_d H_d\right),~\mathrm{and} \\ 
    &V_D = \frac{g_1^2 + g_2^2 }{8} \left(  H^\dagger_u H_u  -  H^\dagger_d H_d\right)^2 + \frac{g_2^2}{2} |H^\dagger_d H_u|^2.
\end{split}    
\label{Eqn:D_term}
\end{equation}
In Eq.~\eqref{Eqn:D_term}, $g_{1}$ and $g_{2}$ are the SM $U(1)_{Y}$ and $SU(2)_{L}$ gauge couplings, respectively. Expanding $V_{Higgs}$ around the real neutral $vev$s $v_d, v_u {\rm ~and~} v_S$, we obtain the physical neutral Higgs fields $\{H^0_d, H^0_u,S\}$,
\begin{equation}
H^0_d = v_d+\frac{H_{dR} + iH_{dI}}{\sqrt{2}},\quad H^0_u = v_u+ \frac{H_{uR} + iH_{uI}}{\sqrt{2}},\quad S = v_S + \frac{S_R + iS_I}{\sqrt{2}},
\label{eqn:Higgs_fields}
\end{equation}
where, the subscripts $R$ and $I$ indicate the CP-even and CP-odd states of $\{H^0_d, H^0_u,S\}$, respectively. In the basis $\{H_{dR}, H_{uR}, S_R\}$, the $3\times3$ symmetric mass-squared matrix $\mathcal{M_S}^{2}$ for the CP-even neutral Higgs states is given by~\cite{Ellwanger:2009dp},
\begin{equation}
\resizebox{1.0\hsize}{!}{$
\mathcal{M_S}^2 = 
\begin{pmatrix}
g^2v_d^2 + \mu(A_\lambda + \kappa v_S)\tan\beta& (2\lambda^2 - g^2)v_uv_d - \mu(A_\lambda + \kappa v_S)& \lambda(2\mu v_d -(A_\lambda + 2\kappa v_S)v_u)\\
 ... & g^2v_u^2 + \mu (A_\lambda + \kappa v_S)/\tan\beta & \lambda(2\mu v_u - (A_\lambda + 2 \kappa v_S) v_d)\\
 ...  & ... & \lambda A_\lambda \frac{v_u v_d}{v_S} + \kappa v_S(A_\kappa + 4 \kappa v_S)\\
\end{pmatrix},
\label{eq:mass_mat_nu}
$}
\end{equation}
where $g^2 = (g_1^2 + g_2^2)/2$, $v_u = v\sin\beta$ and $v_d=v\cos\beta$ with $v^2 = v_u^2 + v_d^2\approx (174~{\rm GeV})^2$, and $\tan\beta = v_u/v_d$. The CP-even Higgs mass eigenstates $H_i~(i=1,2,3)$ can be obtained by diagonalizing $\mathcal{M_S}^2$ by the matrix $S$, such that
\begin{equation}
    H_{i} = \sum_{j=1}^{3} S_{ij}H_{jR}, \quad {\rm with}\quad H_{jR}=\{H_{uR}, H_{dR},S_{R}\}.
\end{equation}
Assuming negligible mixing between the doublet and singlet components, the squared mass of the singlet-dominated CP-even state at the tree-level is given by the (3,3) element of $\mathcal{M_S}^2$,
\begin{equation}
m_{H_S}^2 \approx \mathcal{M_{S}}_{33}^2 = \lambda A_\lambda \frac{v_u v_d}{v_S} + \kappa v_S (A_\kappa + 4 \kappa v_S).
\label{eq:m_singlet_even}
\end{equation}
Under this assumption, one of the other two CP-even eigenstates must be consistent with the observed SM-like Higgs boson with a mass near $125$~GeV, and the other one is a MSSM-like Higgs eigenstate with squared mass around
\begin{equation}
    m_{H_3}^2 = 2\mu (A_\lambda + \kappa v_S)/\sin2\beta
    \label{eq:m_H3}
\end{equation}

In this scenario, the CP-even Higgs bosons can be conveniently rotated to the basis $\{\widehat{h},\widehat{H},\widehat{s}\}$~\cite{Badziak:2015exr,Miller:2003ay}, such that $\widehat{h}=H_{dR}\cos\beta + H_{uR}\sin\beta$, $\widehat{H}=H_{dR}\sin\beta - H_{uR}\cos\beta$ and $\widehat{s}=S_R$, with $\widehat{h}$ and $\widehat{H}$ fields resembling the SM-like Higgs boson and the MSSM-like heavy Higgs bosons, respectively. The mass eigenstates in this basis can then be expressed as, 

\begin{equation}
   H_i =V_{h_i\widehat{h}}\widehat{h} + V_{h_i\widehat{H}}\widehat{H} + V_{h_i\widehat{s}}\widehat{s} \quad  \text{where} ~(i = 1,2,3) \\
\end{equation}

where $V$ represents the $3\times 3$ unitary matrix that diagonalizes the mass-squared matrix of CP-even Higgs bosons in the rotated basis $\{\widehat{h},\widehat{H},\widehat{s}\}$.

The mass of SM-like Higgs boson $H_{SM}$ at the one-loop level can be written as~\cite{Ellwanger:2011sk},
\begin{equation}
     m_{H_{SM}}^2=m_Z^2\cos^22\beta + \lambda^2v^2\sin^22\beta + \Delta_{mix} 
    +\Delta_{\rm rad. corrs.}
    \label{eq:higgs_mass}
\end{equation}
Here, the second term implies that a large $\lambda$ and low $\tan\beta$ is typically favorable to uplift $m_{H_{SM}}$ from $\sim m_{Z}$ and bring it closer to the observed mass of $\sim 125~$GeV at the tree-level, thus typically reducing the amount of radiative corrections required to generate the correct Higgs boson mass, when compared to MSSM. The third term in Eq.~\eqref{eq:higgs_mass}, $\Delta_{\rm mix}$, is the contribution arising from singlet-doublet mixing. In the limit where the mixing is not large, $\Delta_{\rm mix}$ is given by~\cite{Ellwanger:2011sk},  
\begin{equation}
     \Delta_{\rm mix} \simeq \frac{4 \lambda^2 v_S^2 v^2(\lambda - \kappa \sin2\beta)^2}{(\overline{M_h}^2 - M_{SS}^2)}, 
     \end{equation} 
where, $\overline{M_h}^2 = m_{H_{SM}}^2$ (given by Eq.~\eqref{eq:higgs_mass}) for $\Delta_{\rm mix} \sim 0$, and $M_{SS}^2\simeq \kappa v_S(A_\kappa + 4 \kappa v_S)$ is the squared mass of CP-even singlet-like Higgs assuming a heavy singlet, $\kappa v_S >> A_{\kappa}, A_{\lambda}$ (see Eq.~\eqref{eq:m_singlet_even}). The contribution from this term can be positive or negative depending on the mass difference $\overline{M_h}^2 - M_{SS}^2$.

In the decoupling limit, $\lambda,\kappa \to 0 $, the radiative corrections to $m_{H_{SM}}$, $\Delta_{\rm rad. corrs.}$, become crucial to obtain the correct mass for the SM-like Higgs boson. The dominant radiative contributions at one-loop from the top/stop loops are given by, 
\begin{equation}
    \Delta_{\rm rad. corrs.}^{\rm 1-loop} \simeq \frac{3m_{t}^4}{4\pi^2v^2\sin^2\beta}\left[ln\frac{M_{S}^2}{m_t^2} + \frac{X_t^2}{M_S^2}\left(1 - \frac{X_t^2}{12M_S^2}\right)\right],
\end{equation}
where, $m_t$ denotes top quark mass, $M_{S}=\sqrt{m_{\tilde{t}_1} m_{\tilde{t}_2}}$ with $m_{\tilde{t}_1,\tilde{t}_2}$ stands for mass ordered stop squarks and $\left|X_t\right|=A_t -\mu \cot\beta$ is stop mixing parameter with $A_t$ being the soft supersymmetry-breaking stop trilinear coupling. 

Similar to Eq.~\eqref{eq:mass_mat_nu}, the mass squared matrix $\mathcal{M^\prime_{P}}^2$ for the CP-odd states ($H_{dI}, H_{uI}, S_{I}$) from Eq.~\eqref{eqn:Higgs_fields} can be defined as~\cite{Ellwanger:2009dp},
\begin{equation}
   \mathcal{M^\prime_{P}}^2 =
\begin{pmatrix}
    \mu (A_\lambda + \kappa v_S)\tan\beta & \mu (A_\lambda + \kappa v_S) & \lambda v_u(A_\lambda - 2\kappa v_S)\\
    ... & \mu (A_\lambda + \kappa v_S)/\tan\beta & \lambda v_d(A_\lambda - 2\kappa v_S)\\ 
    ... & ... & \lambda(A_\lambda + 4\kappa v_S)\frac{v_u v_d}{v_S} - 3 \kappa A_\kappa v_S\\
\end{pmatrix}.
\end{equation}
Without any loss of generality, the mass matrix $\mathcal{M^\prime_{P}}^2$ can be rotated to a new basis, $\{H_{dI}, H_{uI}, S_{I}\} \to \{A,G,S_{I}\}$, 
\begin{equation}
\begin{pmatrix}
H_{dI}\\
H_{uI}\\
S_{I}
\end{pmatrix}=
\begin{pmatrix}
    \sin\beta & -\cos\beta & 0\\
    \cos\beta & \sin\beta & 0\\
    0 & 0 & 1
\end{pmatrix}
\begin{pmatrix}
    A\\
    G\\
    S_{I}
\end{pmatrix}.
\end{equation}
Dropping the Goldstone mode $G$, the remaining $2 \times 2$ mass squared matrix for CP-odd states in the basis of $\{A,S_I\}$ can be expressed as,

\begin{equation}
   \mathcal{M_{P}}^2 =
\begin{pmatrix}
    m_{A}^2 & \qquad \lambda (A_\lambda - 2\kappa v_S)v\\
    \lambda(A_\lambda - 2\kappa v_S)v & \qquad \lambda(A_\lambda + 4\kappa v_S)\frac{v_u v_d}{v_S} - 3 \kappa A_\kappa v_S\\
\end{pmatrix},
\label{eqn:CP_odd_2x2_matrix}
\end{equation}
with $m_{A}^2 = 2\mu (A_\lambda + \kappa v_S)/\sin2\beta$, which represents the squared mass of the doublet-like CP-odd scalar, $A$, similar to that in the MSSM. Likewise, the squared mass of the singlet-like CP-odd scalar $A_s$, assuming negligible singlet-doublet mixing, is represented by the $\{2,2\}$ element in $\mathcal{M_{P}}^{2}$. 

 Overall, the scalar sector is characterized by six parameters:
 \begin{equation}
 \lambda,\kappa, A_\lambda, A_\kappa, \mu, \tan\beta,
 \end{equation}
 as indicated by Eqs.\eqref{eq:m_singlet_even}, \eqref{eq:higgs_mass} and \eqref{eqn:CP_odd_2x2_matrix}.

The electroweakino sector of NMSSM consists of 5 neutralinos and 2 charginos. The $5\times5$ symmetric neutralino mass matrix $\mathcal{M}_{\widetilde{N}}$ in the basis of $\{\widetilde{B}$: Bino, $\widetilde{W}^0_3$: Wino, $\widetilde{H}_d^0,\widetilde{H}^0_u$: Higgsinos, and $\widetilde{S}$: Singlino$\}$ is given by,

\begin{equation}
    \mathcal{M}_{\widetilde{N}}=
    \begin{pmatrix}
        M_1 & 0 & -\frac{g_1 v_d}{\sqrt{2}} & \frac{g_1 v_u}{\sqrt{2}} & 0 \\
        ... & M_2 & \frac{g_2 v_d}{\sqrt{2}} & -\frac{g_2 v_u}{\sqrt{2}} & 0 \\
        ... & ... & 0 & -\mu_{eff} & - \lambda v_u\\
        ... & ... & ... & 0 & -\lambda v_d\\
        ... & ... & ... & ... &-2\kappa v_S\\
    \end{pmatrix}
    \label{eq:mat_neu}
\end{equation}
where $M_1$ and $M_2$ are the Bino and Wino mass parameters.
The neutralino mass eigenstates $\mlspi$(i=1,..5) can be obtained by diagonalizing $\mathcal{M_{\widetilde{N}}}$ through an orthogonal real matrix $N$, 
\begin{equation}
    \lspi = N_{i1}\widetilde{B} + N_{i2}\widetilde{W}^0_3+ N_{i3}\widetilde{H}_d^0 + N_{i4}\widetilde{H}^0_u + N_{i5}\widetilde{S}; 
    \label{eqn:neutralino_admixture}
\end{equation}
where the neutralino masses are not necessarily positive but ordered according to their absolute value.  

The remaining components in the electroweakino sector are the two charginos~($\tilde{\chi}_i^\pm (i=1,2)$), which are formed through mixing between the charged Winos and Higgsinos. In the basis $\psi^+=(-i\widetilde{W}^+, \widetilde{H}_u^+)$ and $\psi^-=(-i\widetilde{W}^-, \widetilde{H}_d^-)$, the $2\times 2$ asymmetric chargino mass matrix can be written as~\cite{Ellwanger:2009dp},
\begin{equation}
    \mathcal{M_{C}}=
    \begin{pmatrix}
        M_2 & g_2v_u\\
        g_2v_d & \mu
    \end{pmatrix}
\end{equation}
The diagonalization of $\mathcal{M_C}$ is accomplished by two different rotations, $\psi^+$ and $\psi^-$ denoted by two unitary matrix $U$ and $V$, respectively:
\begin{equation}
    U^* \mathcal{M_C}V^\dagger = diag(m_{\chonepm}, m_{\chtwopm}); \quad {\rm with~}m_{\chonepm} < m_{\chtwopm}.
\end{equation}
Overall, the electroweakino sector at the tree-level is characterized by six input parameters: 
\begin{equation}
M_1,M_2,\mu ,\tan\beta,\lambda,\kappa.
\end{equation}

In the R-parity conserved NMSSM, the lightest neutralino $\lspone$ emerges as a potential candidate for dark matter. As indicated by Eq.~\eqref{eqn:neutralino_admixture}, LSP $\lspone$ can be either $\widetilde{B}$, $\widetilde{W}$, $\widetilde{H}$ or $\widetilde{S}$-like, or a hybrid of these states. While the pure $\widetilde{B}$ and $\widetilde{S}$-like $\lspone$ results in a correct relic abundance through resonant annihilation or co-annihilation, Winos or Higgsinos up to masses of 2.8 TeV tends to produce the correct relic density through self-annihilation. 


\section{Parameter space scan and constraints}
\label{sec:constraints}
In this work, we concentrate on the region of $Z_3$-invariant NMSSM parameter space that leads to Singlino-dominated LSP, $\lspone$, with under-abundant relic density. In the $\lambda \to 0$ limit, the singlet superfield decouples from the MSSM Higgs superfields, as indicated by Eq.~\eqref{eq:superpot}. In this decoupled regime, the mass of the Singlino-like neutralino at the tree-level is approximately 
$\sim 2 \kappa v_S$. Given the effective Higgsino mass parameter $\mu = \lambda v_S$, and our region of interest being the parameter space with a Singlino-dominated LSP $\lspone$, we require $2 \kappa v_S \lesssim \lambda v_S$. Accordingly, we impose $|\kappa/\lambda| < 0.5$ in our scans to ensure Singlino dominance in the LSP $\lspone$. Values of $\kappa, \lambda < 10^{-3}$ are avoided as to ensure beyond-the-MSSM scenario. It is also worth noting that perturbativity of the theory up to the GUT scale requires both $\lambda,\kappa < 0.7$. Based on initial results from our scan, we restrict $\lambda$ to less than 0.4 to increase the likelihood of generating points with Singlino-dominated LSP. Our scan considers a wide range for the Bino, Wino and Higgsino mass parameters: $70~\mathrm{GeV}< M_1 < 2~\mathrm{TeV}$, $200~\mathrm{GeV}< M_2 < 3~\mathrm{TeV}$ and $100~\mathrm{GeV}< \mu < 900~\mathrm{GeV}$, respectively. $M_2$ is varied from a higher minimum value due to more stricter constraints on Winos. This choice of $M_1$, $M_2$, and $\mu$ allows the study of a diverse electroweakino sector within the NMSSM, involving various admixtures from Bino, Wino, and Higgsinos in the Singlino-dominated LSP. A small value of $A_{\kappa}$ tends to favor a singlet-dominated light Higgs bosons, which could be crucial for resonant DM annihilation at low LSP masses, $m_{\lspone} \lesssim m_{Z}/2$. Therefore, we vary it in the [0-1000]~GeV range. Other key input parameters required to characterize the Higgs and electroweakino sectors at the tree level are $A_{\kappa}$ and $\tan\beta$, which have been varied in the [-10,$10^4$]~GeV and [1,40] ranges, respectively. The third-generation squark mass parameters $MQ_3, MU_3, MD_3$ are varied in the range  [500-$10^{4}$]~GeV, while the third-generation slepton mass parameters are fixed at 3~TeV. The second-generation squark and slepton mass parameters are also fixed at 3~TeV. Lastly, we vary the stop trilinear coupling $A_t$ in the range [-10,10] TeV, while the bottom and stau trilinear couplings are fixed at $A_b = A_\tau =2$ TeV. 

We perform a random scan over the parameter space, considering the previously discussed scan range, utilizing the \texttt{NMSSMTools-v6.0.0}~\cite{Ellwanger:2004xm, ELLWANGER2006290, Das:2011dg} package. The masses and relevant branching ratios of the Higgs bosons and the SUSY particles are computed using \texttt{NMSSMTools-v6.0.0}~\cite{Ellwanger:2004xm, ELLWANGER2006290, Das:2011dg}. We use \texttt{MicrOmegas}~\cite{Belanger:2006is, Belanger:2008sj, Barducci:2016pcb} to compute the observables constrained by LEP, $B$-physics experiments, and DM measurements. We initially perform a flat random scan over $10^8$ points. Roughly $10^{-4}\%$ of the scanned points pass the relevant collider and astrophysical constraints. The range of input parameters considered in the scan is summarized below: 

\begin{equation}
\begin{split}
& 0.0001 < \lambda < 0.4,~\left|\frac{\kappa}{\lambda}\right|\leq 0.5,~M_1 = (70,2000)~{\rm GeV},\\
&M_2=(200,3000)~{\rm GeV},~M_3= (2000, 5000)~{\rm GeV},\\
&\mu= (100,900)~{\rm GeV},~\tan\beta = (1,40),\\
&A_\lambda = (-10,10000)~{\rm GeV},~|A_\kappa| = (0,1000)~{\rm GeV},\\
&A_{t} = (-10000, 10000)~\mathrm{GeV}\, \\ 
&A_{b} = A_{\tau}=2000\mathrm{~GeV} \\
&ML_3=ME_3=2000 \mathrm{~GeV}\\
&MQ_3=MU_3=MD_3 =(0.5,10) \mathrm{~TeV}
\label{eq:param_scan}
\end{split}
\end{equation}

\subsection*{Constraints from collider experiments}

\begin{figure*}[!t]
	\centering
	\includegraphics[scale=0.45]{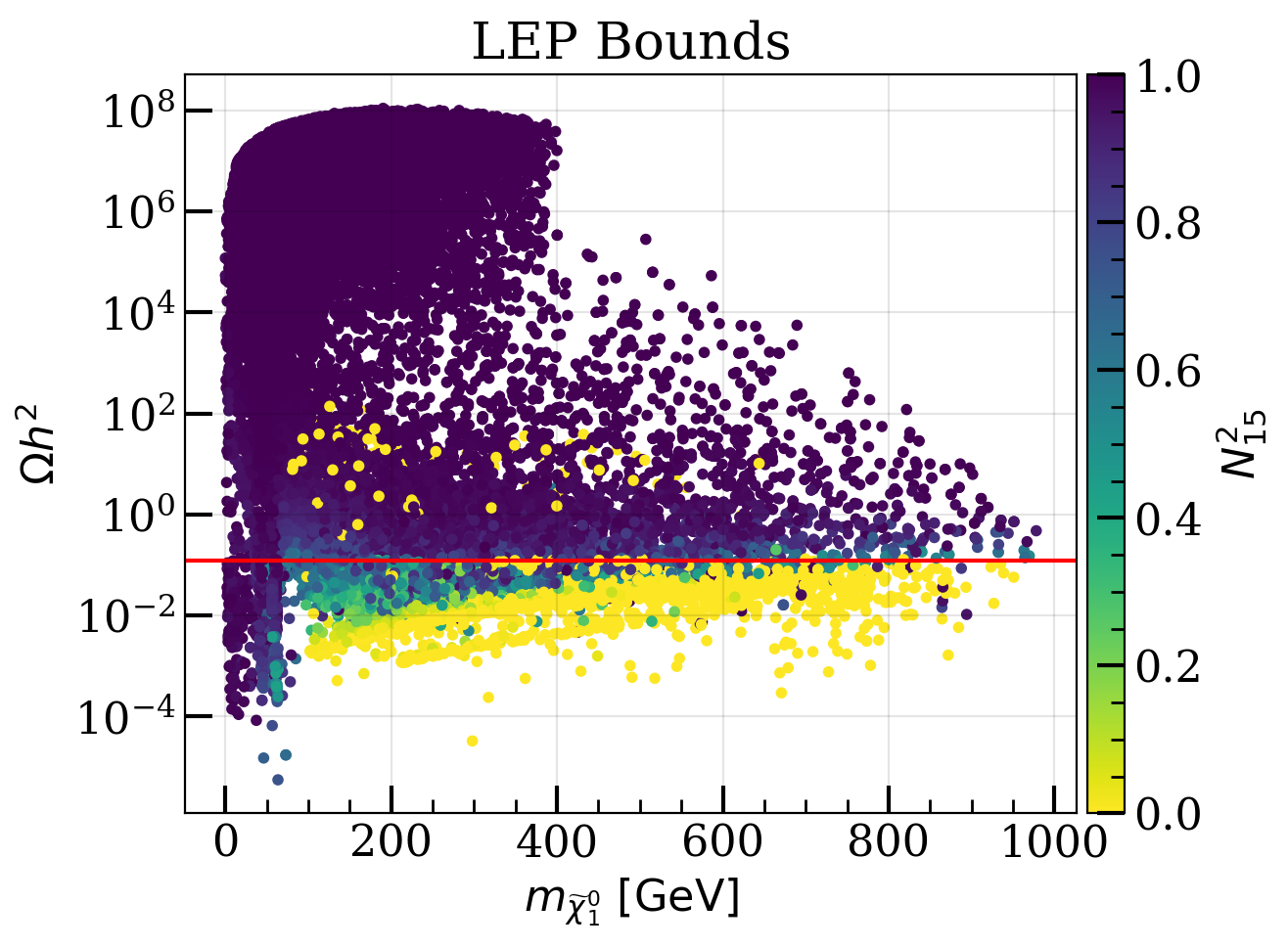}\hfill
	\includegraphics[scale=0.45]{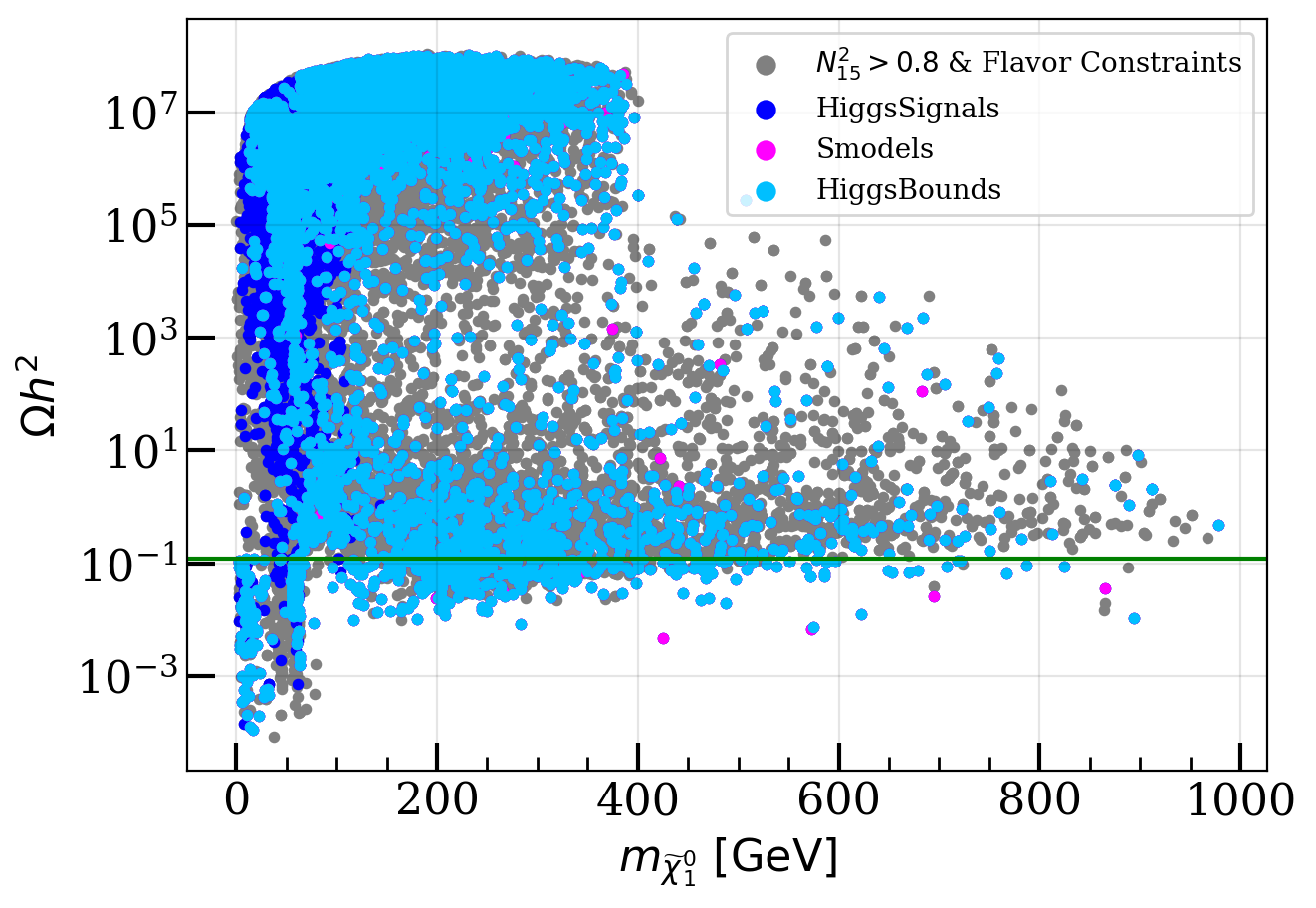} 
	\caption{\textit{Left:} Parameter points allowed by the theoretical constraints and LEP limits are shown in the $m_{\lspone}$-$\Omega h^{2}$ plane. The color palette represents the Singlino admixture $N_{15}^2$ in the LSP $\lspone$. \textit{Right:} Parameter points allowed by the successive application of flavor constraints and the requirement of a Singlino-dominated $\lspone$~($N_{15}^{2} > 0.8$)~(shown in grey), Higgs signal-strength constraints imposed through \texttt{HiggsSignals-v2.6.2}~\cite{Bechtle:2013xfa, Bechtle:2014ewa}~(blue), bounds from electroweakino, gluino, squark, stop and sbottom searches at the LHC applied through \texttt{SModels-v2.3.2} \cite{Alguero:2021dig, MahdiAltakach:2023bdn}~(pink) and the limits from BSM Higgs searches at the LHC implemented via \texttt{HiggsBounds-v5.10.0}~\cite{Bechtle:2008jh, Bechtle:2011sb, Bechtle:2013wla, Bechtle:2013xfa}~(skyblue), are shown in the same plane.}
\label{fig:const}
\end{figure*}
The NMSSM parameter space of our interest is constrained by various collider and astrophysical measurements. We first discuss the relevant collider constraints. Adhering to LEP measurements, we require the mass of the charginos to be $M_{\chonepm} > 103.5~$GeV~\cite{OPAL:2003wxm} and the production rate of the process $e^+e^- \to \lsptwo\lspone$ in the jets+$\met$ final state with $|M_{\lsptwo} - M_{\lspone}| < 5~$GeV to be below $\leq 0.1~$pb at $95\%$ C.L. Limits from searches in the $Z H_j$ and $A_i H_j$ processes are imposed through \texttt{NMSSMTools-v6.0.0} \cite{Ellwanger:2004xm, ELLWANGER2006290, Das:2011dg}. In the left panel of Fig.~\ref{fig:const}, we display the parameter space in the $m_{\lspone}$-$\Omega h^2$ plane, consistent with the LEP limits. The color axis represents the amount of Singlino admixture, $N_{15}^2$ in $\lspone$. The imposed upper limit on the relic density $\Omega_{\lspone}h^2 < 0.122$ is indicated as a solid red line in Fig.~\ref{fig:const}~(left).
Furthermore, for these over-abundant points, it is typically observed that $|\kappa|/\lambda > 0.3$ in the higher $m_{\lspone} \gtrsim 100$ region. Such points typically lead to an MSSM-like scenario with Higgsino-like dark matter. Notably, around $24\%$ of points below $\Omega_{\lspone}h^2 < 0.122$ are Singlino-domimated with $N_{15}^2 > 0.8$, which are the primary focus on this analysis.  

The parameter space of interest is also constrained by flavor physics observables, especially through measurements of the rare decay of $B$-mesons. Updated constraints from flavor physics are also imposed from \texttt{NMSSMTools-v6.0.0} \cite{Ellwanger:2004xm, ELLWANGER2006290, Das:2011dg} in terms of 
$BR(b \to s \gamma),~ BR(B_s \to \mu^+\mu^-),~BR(B^+ \to \tau^+ \nu_\tau),~BR(B \to X_s \mu^ + \mu^-)$. Additionally constraints from $\Upsilon (1s) \to H/ A \gamma$, $\Delta M_s, \Delta M_d, \eta_b (1s)$ mass difference are imposed.

Among the three CP-even Higgs bosons, we require either $H_1$ or $H_2$ to be consistent with the properties of the observed SM-like Higgs boson $H_{SM}$. The mass of $H_{SM}$ has been measured at $125.28 \pm 0.14~$GeV through combined measurements by the ATLAS and CMS collaborations at the LHC \cite{ATLAS:2012yve, CMS:2012qbp}. To account for the theoretical uncertainties in the Higgs boson mass calculation~\cite{ATLAS:2015yey}, we adopt a conservative approach and require the mass of the Higgs boson consistent with $H_{SM}$ to be within 122-128~GeV. Its couplings with $t\bar{t}$, $b\bar{b}$, $\tau^+\tau^-$, $\gamma\gamma$, $W^+W^-$, $ZZ$ and $gg$ are also required to be consistent with the signal strength measurements within $2\sigma$ uncertainty, imposed through the \texttt{HiggsSignals-v2.6.2}~\cite{Bechtle:2013xfa, Bechtle:2014ewa} package. 
The heavier and/or lighter scalar and pseudoscalar Higgs bosons can be an admixture of doublet and singlet states and are primarily constrained by BSM Higgs searches at the LEP, Tevatron and the LHC. We impose these constraints using \texttt{HiggsBounds-v5.10.0}~\cite{Bechtle:2008jh, Bechtle:2011sb, Bechtle:2013wla, Bechtle:2013xfa}. Limits from the searches for supersymmetric particles at the LHC, especially electroweakinos, are implemented using \texttt{SModels-v2.3.2} \cite{Alguero:2021dig, MahdiAltakach:2023bdn} interfaced with \texttt{NMSSMTools-v6.0.0} \cite{Ellwanger:2004xm, ELLWANGER2006290, Das:2011dg}.

\subsection*{Constraints from relic density of dark matter and direct detection} 

As discussed previously, our main focus is the region of parameter space where the Singlino-dominated LSP $\lspone$ serves as the thermal DM candidate. In this scenario, the Dark Matter relic abundance of $\lspone$ is required to be within $\Omega_{\lspone} h^{2} < 0.122$, allowing a $2\sigma$ uncertainty around the best-fit of $\Omega_{DM}^{obs.} h^{2} = 0.120 \pm 0.001$ as measured by the PLANCK collaboration~\cite{Planck:2018vyg}. 

\begin{figure}[!t]
\centering
\includegraphics[scale=0.45]{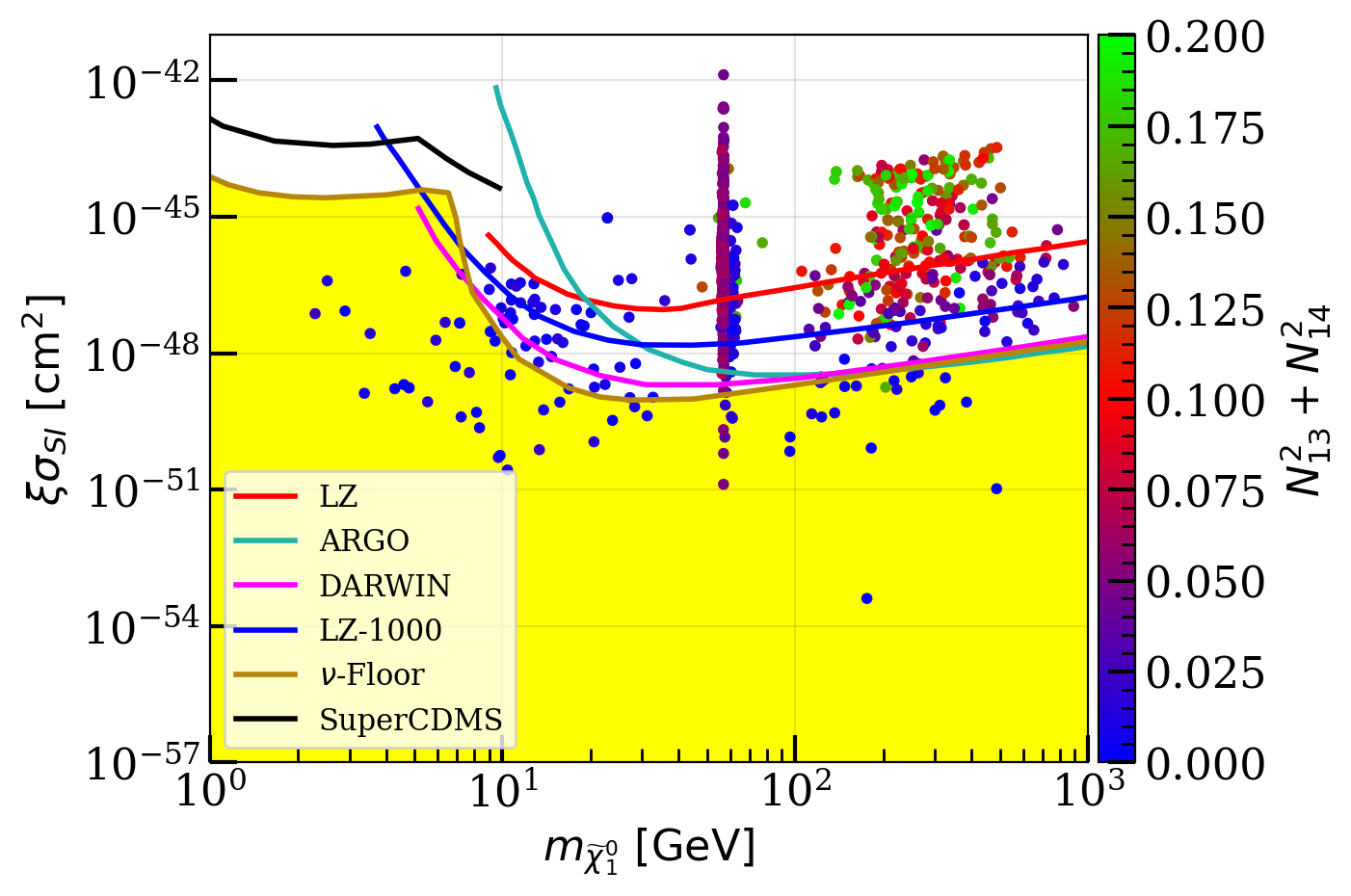}
\caption{Parameter space points with a Singlino-dominated LSP, and allowed by searches at the LEP, rare $B$-meson decay, Higgs signal strength, and BSM Higgs, electroweakino and other particle searches at the LHC~(sky-blue points from the right panel of Fig.~\ref{fig:const}) are shown in the $\xi \sigma_{SI} - m_{\lspone}$ plane. The color palette represents the Higgsino fraction in the LSP. The red lines denote the current upper limit of $\sigma_{SI}$ from the LZ results at 90\% C.L~\cite{PhysRevLett.131.041002}. Projected limits on $\sigma_{SI}$ from future experiments, ARGO~\cite{argo}, DARWIN~\cite{Schumann015:2cpa, DARWIN:2016hyl}, and LZ-1000~\cite{LZ:2018qzl} are also presented.}
	\label{fig:LZ_1}
\end{figure}

The parameter space points that survive the previously discussed constraints are further subjected to the most recent upper limits on the spin-independent DM-nucleon cross-sections $\sigma_{SI}$, from LZ~\cite{PhysRevLett.131.041002}. We also impose the upper limits from SuperCDMS~\cite{SuperCDMS:2016wui}, which is sensitive in the lower $m_{DM}$ regime, $100~\textrm{MeV} \lesssim m_{DM} \lesssim 5~\textrm{GeV}$. We rescale $\sigma_{SI}$ with $\xi$, the ratio of the DM relic density predicted by the model to the observed upper limit on $\Omega_{DM}^{obs.} h^{2}$ allowing $3\sigma$ uncertainty, $\xi = \Omega h^{2}/0.122$. In Fig.~\ref{fig:LZ_1}, we illustrate the parameter points allowed by the previously discussed constraints in the plane of $\xi \sigma_{SI}$ and $m_{\lspone}$. Among them, approximately $70\%$ of points are excluded by LZ SI limits. We also illustrate the projected sensitivity of ARGO~\cite{argo}, DARWIN~\cite{Schumann015:2cpa, DARWIN:2016hyl}, and LZ-1000~\cite{LZ:2018qzl}, in probing $\sigma_{SI}$ in the same plot. It is observed that a sizeable fraction of currently allowed points are within the reach of these future experiments. Several of these currently allowed points also fall beneath the neutrino scattering floor shown in yellow, thus will remain outside the reach of any $\sigma_{SI}$-based future direct detection experiments. 

The spin-independent DM-nucleon interaction is largely mediated through the $t$-channel exchange of CP-even Higgs bosons, where $\sigma_{SI}$ can be expressed as~\cite{Cao:2018rix},
\begin{equation}
    \sigma_{SI} = \frac{4 \mu_R}{\pi} \left| f\right|^2,\quad f \approx \sum_{i=1}^3
    f_{H_i} = \sum_{i=1}^3 \frac{g_{H_i \lspone \lspone} g_{H_i N N}}{2m_{H_i}^2},
    \label{eq:sigmaSI}
\end{equation}
where $N$ denotes the nuclear states, $\mu_R$ is the reduced mass of the DM particle and the nucleon, and $g_{H_i \lspone \lspone}$ and $g_{H_i N N}$ denotes the coupling of the $i^{\text{th}}$ CP-even Higgs bosons with the DM particle and the nucleons, respectively. For a Singlino-dominated $\lspone$, characterized by $N_{15}^2 \sim 1$, the couplings with $H_i$ are given by~\cite{Badziak:2015exr},

\begin{multline}
    g_{H_i \lspone \lspone}\Big|_{\{\widehat{h},\widehat{H},\widehat{s}\}} \approx \sqrt{2}\lambda \big[ V_{H_i \widehat{h}} N_{15}(N_{13} \sin\beta + N_{14}\cos\beta) +  V_{H_i \widehat{H}} N_{15}(N_{14} \sin\beta - N_{13}\cos\beta)  \\
    +  V_{H_i \widehat{s}}(N_{13}N_{14} - \frac{\kappa}{\lambda}N_{15}^2) \big],
    \label{eq:gchichiH}
\end{multline}
where we have retained $N_{15}$ or $V_{H_i\hat{S}}$ dependent terms only. The Singlino-like $\lspone$ couples with the CP-even Higgs bosons through its mixing with the Higgsinos. Additionally, it can couple with the singlet component on its own accord, with the coupling strength proportional to $\kappa/\lambda$, as indicated by Eq.~\eqref{eq:gchichiH}.

The coupling of the $i^{\text{th}}$ CP-even scalar with the nucleons is given by \cite{Badziak:2015nrb}
\begin{equation}
    g_{H_i N N}\Big|_{\{\widehat{h},\widehat{H},\widehat{s}\}} = \frac{m_N}{\sqrt{2}v}\big[V_{H_i\widehat{h}}(F_d^{(N)}+F_u^{(N)})+V_{H_i\widehat{H}}\big(\tan\beta F_d^{(N)}-\frac{1}{\tan\beta}F_u^{(N)}\big)\Big]
    \label{eq:gNNH}
\end{equation}
where $m_N$ denotes the mass of the nucleon and $F_{d,u}^{(N)}$ are the combined form factors associated with the atomic nuclear states. The coupling $g_{H_i N N}$ is primarily determined by the doublet-admixtures in the CP-even Higgs bosons and would be highly suppressed for a singlet-like Higgs boson.  

Overall, for a Singlino-dominated $\lspone$, a higher spin-independent DM-nucleon interaction rate is typically associated with greater Higgsino mixing, as suggested by Eqns.~\eqref{eq:gchichiH} and \eqref{eq:gNNH}. To illustrate this, we show the amount of Higgsino admixture in the LSP $\lspone$, $N^2_{13} + N^2_{14}$, as a color palette in Fig.~\ref{fig:LZ_1}. It is observed that for higher DM masses, $m_{\lspone} \gtrsim 100~$GeV, where consistency with the upper limit on relic density is achieved through co-annihilation~(further discussed in Sec.~\ref{sec:annLSP}), the parameter points excluded by the recent LZ limits are associated with a relatively larger Higgsino admixture. In contrast, in the lower $m_{\lspone}$ regime, a large Higgsino admixture is not required for large $\sigma_{SI}$ due to $s$-channel resonant annihilation mediated through a light Higgs boson.

It is important to note that there could be several conditions that result in a diminished value of $\sigma_{SI}$, falling below the neutrino floor. This could potentially arise from ``blind spots'' in the parameter space~\cite{Badziak:2015exr}. One such example is when coupling of SM-like Higgs with the Singlino-like $\lspone$'s, $g_{H_{SM} \lspone \lspone}$, vanishes for either small lambda or $m_{\lspone}/\mu \simeq \sin 2\beta$~\cite{Meng:2024lmi}. This SM-Higgs exchange process only applies when other CP-even Higgs bosons, $H_s ~\text{and}~H$ are decoupled and $H_s >> H_{SM}$. Another important case related to our present study is the possibility of destructive interference from the CP-even singlet-like Higgs and the SM-like Higgs, $\sigma_{SI} \propto \left[ V_{SM,\widehat{h}}~g_{H_{SM} \lspone \lspone}/m_{H_{SM}}^2 + 
V_{SM,\widehat{S}}~ g_{H_s \lspone \lspone}/m_{H_{s}}^2 \right]^2$. This is critical for $m_{H_s} < m_{H_{SM}}$ with the MSSM like Higgs boson, $H$ being decoupled for both cases of Singlino-dominated LSP having large $\lambda$ and small $\tan\beta$ and Higgsino-Singlino scenario. We refer the reader to Ref.~\cite{Badziak:2015nrb} for further details. Also, it must be noted that a relatively light Bino/Wino-like state, characterized by $m_{\lspone} < M_1 < \mu < M_2$ or $m_{\lspone} < M_1 < M_2 < \mu $, can induce gaugino-admixtures (gaugino is referred to Bino and Wino combinedly) in the Singlino-dominated LSP, which can potentially modify the usual blind spot condition for $Z_3$ symmetric NMSSM, revealing new parameter space for low $\sigma_{SI}$~\cite{Abdallah:2019znp, Roy:2024yoh}.
\begin{figure}[!t]
	\centering
	\includegraphics[scale=0.43]{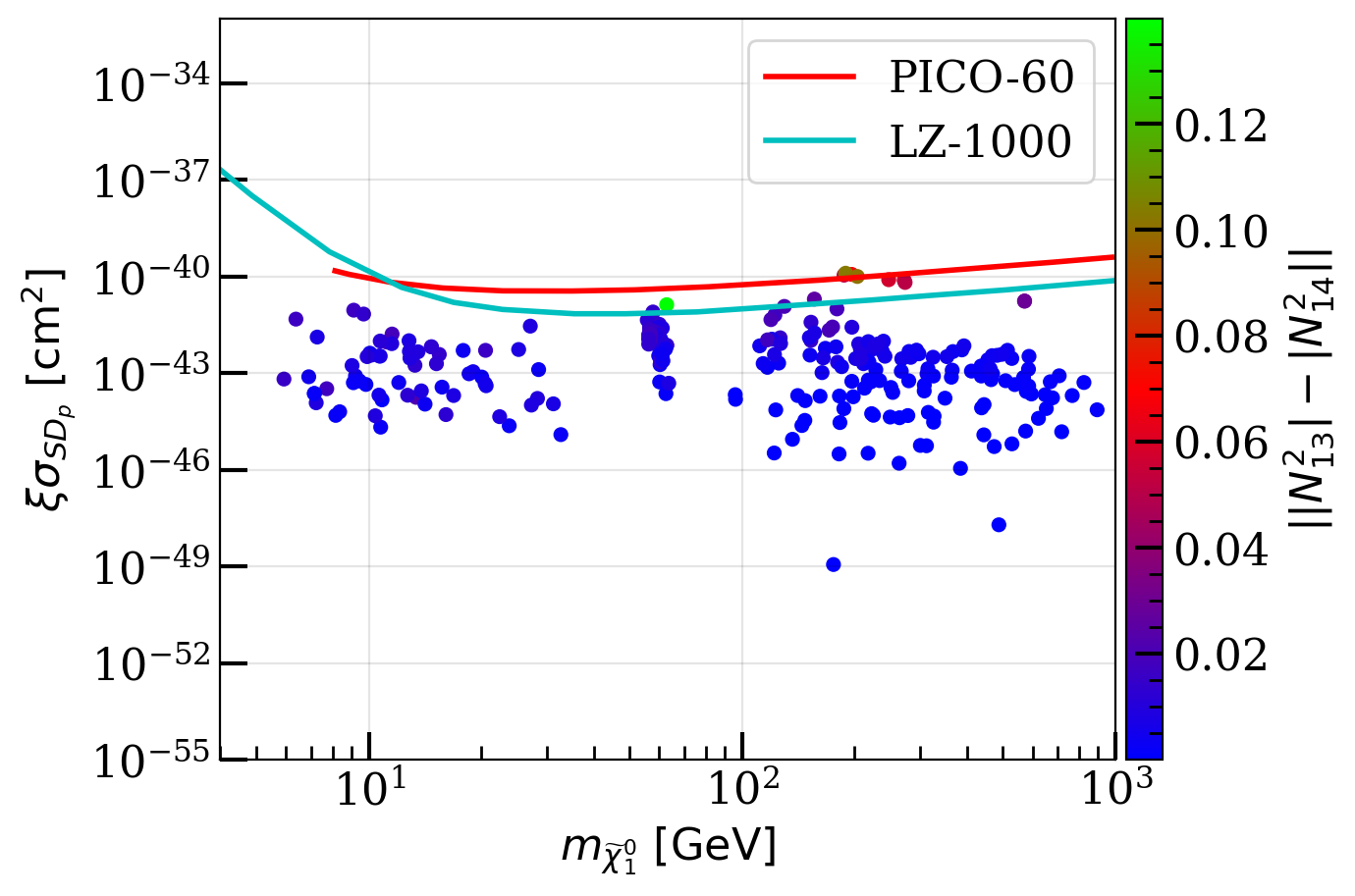}\hfill
	\includegraphics[scale=0.43]{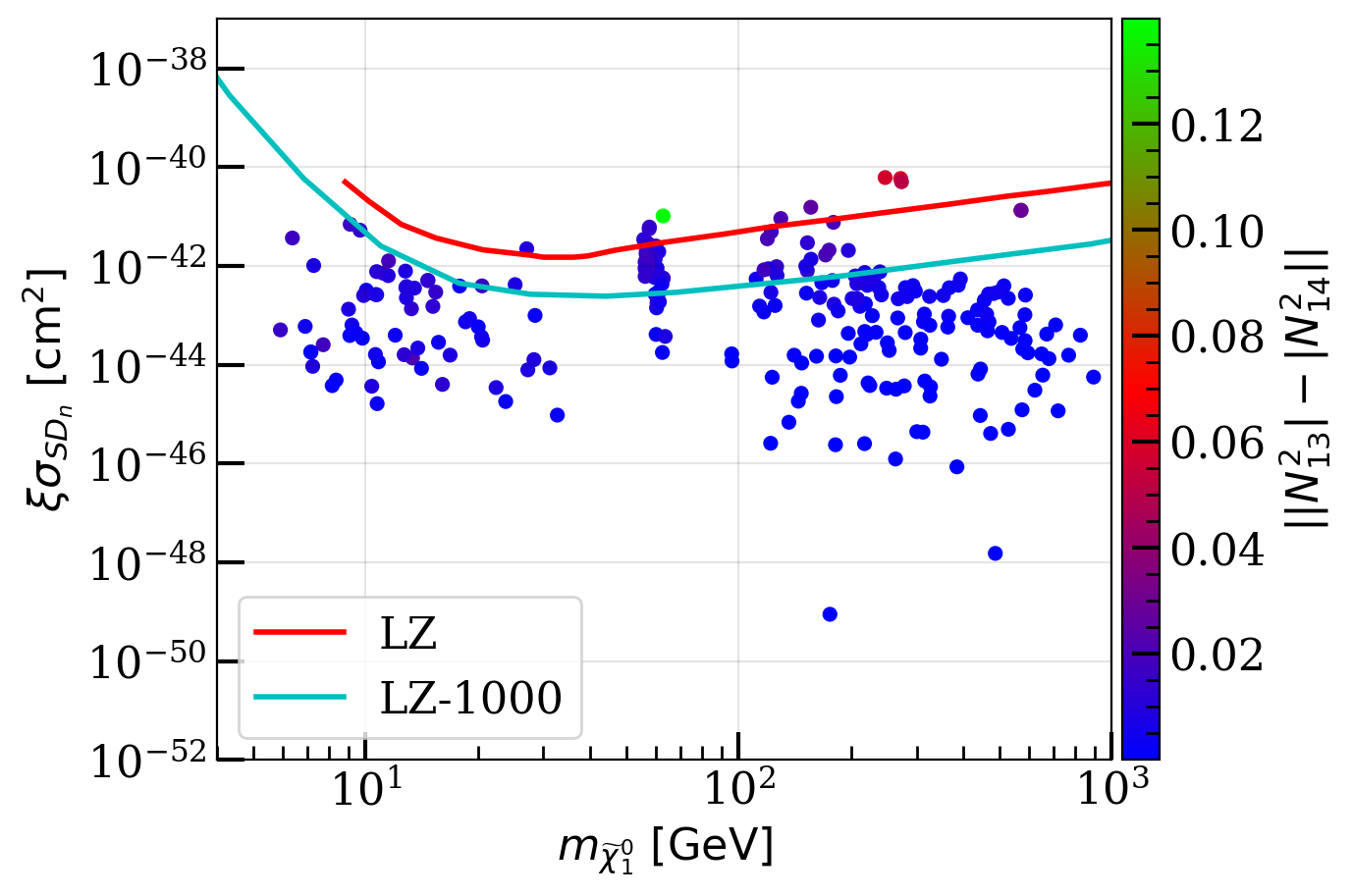}
    \caption{ {\it Left}: The parameter space points allowed by the latest LZ SI limits in Fig.\ref{fig:LZ_1} are plotted in the $\xi \sigma_{SD_p} - m_{\lspone}$ plane. The red line represents the latest upper limit on $\sigma_{SD_p}$ from PICO-60 \cite{PhysRevD.100.022001}. {\it Right}: The allowed parameter points from the {\it Left} pannel are plotted in the $\xi \sigma_{SD_n} - m_{\lspone}$ plane. The current upper limit on $\sigma_{SD_n}$ from LZ \cite{PhysRevLett.131.041002} are presented in red. The future sensitivity from LZ-1000 \cite{LZ:2018qzl} on $SD_p$ and $SD_n$ is also projected in both panels. The color palette in both the panels represent the coupling $g_{Z \lspone \lspone} \propto N_{13}^2 - N_{14}^2$.}
	\label{fig:LZ_2}
\end{figure}

The parameter space of interest can also be probed through spin-dependent DM-nucleon interactions. In this regard, we impose the most recent upper limits on the spin-dependent DM-neutron $\sigma_{SD_n}$ and DM-proton $\sigma_{SD_p}$ cross-sections from LZ~\cite{PhysRevLett.131.041002} and PICO-60~\cite{PhysRevD.100.022001}, respectively, on the parameter points allowed by the previously discussed constraints, including the upper limits on $\xi\sigma_{SI}$. We show our results in Fig.~\ref{fig:LZ_2}. It is observed that only a few parameter space points are excluded by the upper limits on the spin-dependent interaction rate. Future projections for $\sigma_{SD_n}$ and $\sigma_{SD_p}$ for LZ-1000~\cite{LZ:2018qzl} are also shown in the respective figures. The points allowed by the current upper limits on SD DM-neutron and DM-proton interactions will be henceforth referred to as the currently allowed parameter points.  

The DM SD interactions are typically induced by $t$-channel exchange of $Z$ boson. Their cross-section can be expressed as,
\begin{equation}
    \sigma_{SD_{n/p}} \simeq C^{n/p} \times \left(\frac{g_{Z \lspone \lspone}}{0.01} \right)^2,
    \label{eq:SD_gZchichi}
\end{equation}
where n~(p) denotes neutron~(proton), with the nuclear form factor $C^n~(C^p) \sim 10 ^{-41}~\mathrm{cm}^2$, and the coupling $g_{Z\lspone\lspone}$ is determined by the Higgsino admixtures in the LSP $\lspone$,
\begin{equation}
    g_{Z\lspone\lspone}  = \frac{m_Z}{\sqrt{2}v} (N_{13}^2 - N_{14}^2).
\label{eq:coup_gzlsplsp}
\end{equation}
We illustrate the dependency of DM SD cross-sections on $||N_{13}|^2 - |N_{14}^2||$ in Fig.~\ref{fig:LZ_2} through the color palette. It is observed that the parameter points excluded by PICO-60 and LZ, through upper limits on $\sigma_{SD-p}$ and $\sigma_{SD-n}$, respectively, are mostly in red, which corresponds to a relatively larger $||N_{13}|^2 - |N_{14}^2||$. We would like to mention that all the allowed parameter points shown in Fig.~\ref{fig:LZ_2} survive the constraints from DM indirect detection experiments~\cite{Fermi-LAT:2016uux}, as shown in Appendix~\ref{sec:appendixA} .  

\section{Annihilation processes with Singlino LSP}
\label{sec:annLSP}

In this section, we examine the primary DM annihilation modes at different LSP masses. As discussed previously, our focus in this work is the region of parameter space where the LSP $\lspone$ is Singlino-dominated, is consistent with the upper limit on the relic density $\Omega h^2 < 0.122$, and passes the relevant experimental constraints discussed in Sec.~\ref{sec:constraints}. The parameter space of interest contains points where the mass of the LSP spans from approximately 4~GeV to 1~TeV. Different DM annihilation modes ensure complicity with the relic density constraint at different LSP masses. 

\begin{figure}[!t]
	\centering
        \includegraphics[scale=0.44]{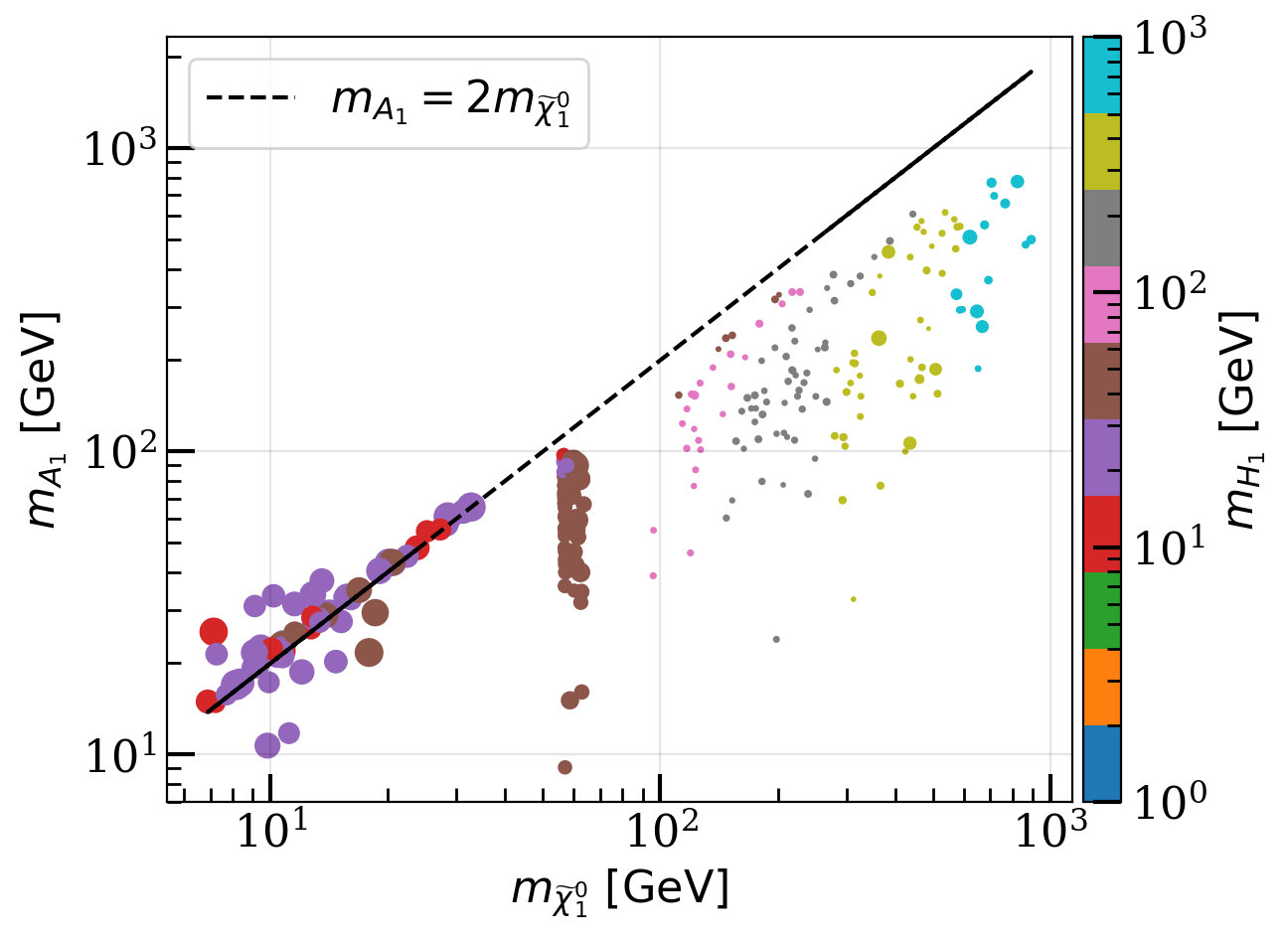}\hfill
	\includegraphics[scale=0.44]{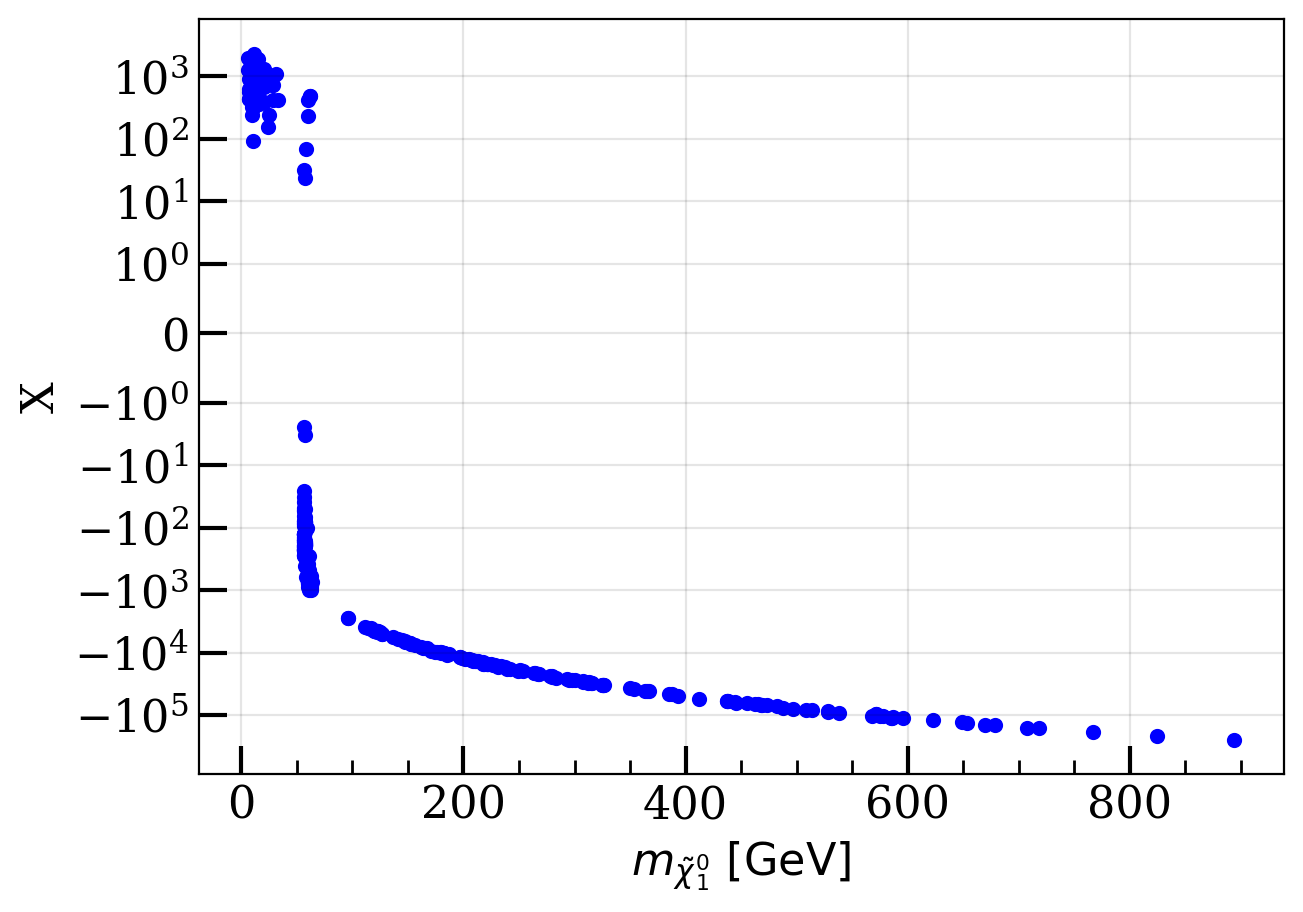}
	\caption{\textit{Left:} The currently allowed parameter points are presented in the plane of $m_{\lspone}$ and the mass of the singlet-dominated pseudoscalar Higgs boson $A_1$. The color palette represents the mass of the singlet-dominated scalar Higgs boson $H_1$. The diagonal black-dashed line corresponds to the condition: $m_{A_1} = 2 m_{\lspone}$. \textit{Right:} The variation of $X$ in Eq.~\eqref{eq:sum_rule_1} with the mass of the Singlino-dominated LSP $\lspone$ is shown for the currently allowed parameter space points. } 
\label{fig:eq_plot}
\end{figure}

Within our parameter space of interest, the DM annihilation mechanisms include resonant $s$-channel annihilation via the exchange of singlet-like Higgs bosons $A_1/H_1$, the $Z$ boson, and the SM-like Higgs boson $H_{SM}$, $t$-channel annihilation via the exchange of a chargino or neutralino, and co-annihilation with the NLSP. For LSP masses below the $Z$-funnel, $m_{\lspone} \lesssim m_{Z}/2$, it is observed that DM annihilation primarily proceeds via the $s$-channel exchange of a singlet-like Higgs boson $A_1/H_1$ with mass $m_{A_1/H_1} \sim 2 m_{\lspone}$. This is depicted in the left panel of Fig.~\ref{fig:eq_plot}, where we show the currently allowed parameter space points in the $m_{\lspone}$-$m_{A_1}$ plane, with the $z$-axis representing $m_{H_1}$. The black-dashed line represents the condition $m_{A_1} = 2m_{\lspone}$, indicating that a major fraction of points align with the criteria. For points in the lower LSP mass regime that are farther away from the $m_{A_1} = 2 m_{\lspone}$ line, it is observed that the lightest CP-even scalar $H_1$ is singlet-like, with mass $m_{H_1} \sim 2 m_{\lspone}$, indicating that DM annihilation proceeds through the exchange of resonant $H_1$ in the $s$-channel. Resonant annihilation predominantly proceeds via the exchange of the $Z$ boson and the SM-like Higgs boson $H_{SM}$ as we approach the $Z$ and $H_{SM}$-funnel thresholds, corresponding to $m_{\lspone} \sim m_{Z}/2$ and $m_{H_{SM}}/2$, respectively.

For the scenario involving a Singlino-like $\lspone$ and the singlet-like Higgs bosons with negligible doublet-admixture, using Eqs.~\eqref{eq:m_singlet_even}, \eqref{eqn:CP_odd_2x2_matrix} and \eqref{eq:mat_neu}, the mass-rule connecting the two can be approximated as~\cite{Ellwanger:2018zxt, Abdallah:2019znp},  
\begin{equation}
\begin{split}
\mathcal{M}_{\widetilde{N},55}^{2} \equiv 4 \kappa^2 v_S^2 = & M_{S_R, S_R}^2 + \frac{1}{3} M_{S_I, S_I}^2 - \frac{4}{3}v_u v_d \left( \lambda^2 \frac{A_\lambda}{\mu} + \kappa\right) \\
& \approx \mathcal{M}_{\mathcal{S},33}^2 + \frac{1}{3} \mathcal{M}_{\mathcal{P},22}^2 - \frac{4}{3}v_u v_d \left( \lambda^2 \frac{A_\lambda}{\mu} + \kappa\right).
\end{split}
\label{eq:sum_rule}
\end{equation}
Considering the singlet-like pseudoscalar Higgs mass twice the mass of the Singlino-like LSP $\lspone$, $\mathcal{M}_{\mathcal{P},22}^2 = 4 \mathcal{M}_{\widetilde{N},55}^{2} = 4 m_{\lspone}^2$, the condition necessary for resonant DM annihilation, Eq.~\eqref{eq:sum_rule} leads to,  
\begin{equation}
\mathcal{M}_{\mathcal{S},33}^2 = -\frac{1}{3} m_{\lspone}^2 + \frac{4}{3} v_u v_d\left(\lambda^2 \frac{A_\lambda}{\mu} + \kappa \right) \equiv X. 
\label{eq:sum_rule_1}
\end{equation}
We illustrate the variation of $X$ in Eq.~\eqref{eq:sum_rule_1} with the mass of Singlino-like LSP in the right panel of Fig.~\ref{fig:eq_plot}. Within the parameter space of our interest, $X$ attains positive values only in the region $m_{\lspone} \lesssim 50~$GeV. Eq.~\eqref{eq:sum_rule_1} and Fig.~\ref{fig:eq_plot}~(right) indicates that it is impossible to simultaneously achieve a physical mass for the singlet-like CP-even Higgs boson, $\mathcal{M}_{\mathcal{S},33}^2 > 0$, given $\mathcal{M}_{\mathcal{P},22}^2 \sim m_{A_1}^{2} \geq 4 \mathcal{M}_{\widetilde{N},55}^{2} \sim m_{\lspone}^2$ for the Singlino-like $m_{\lspone} \gtrsim 50~$GeV. This implies that within our parameter space, resonant DM annihilation through the exchange of a singlet-like Higgs boson in the $s$-channel is not viable for LSP masses above $\gtrsim 50~$GeV. 
 
\begin{figure}[!t]
    \centering
    \includegraphics[scale=0.44]{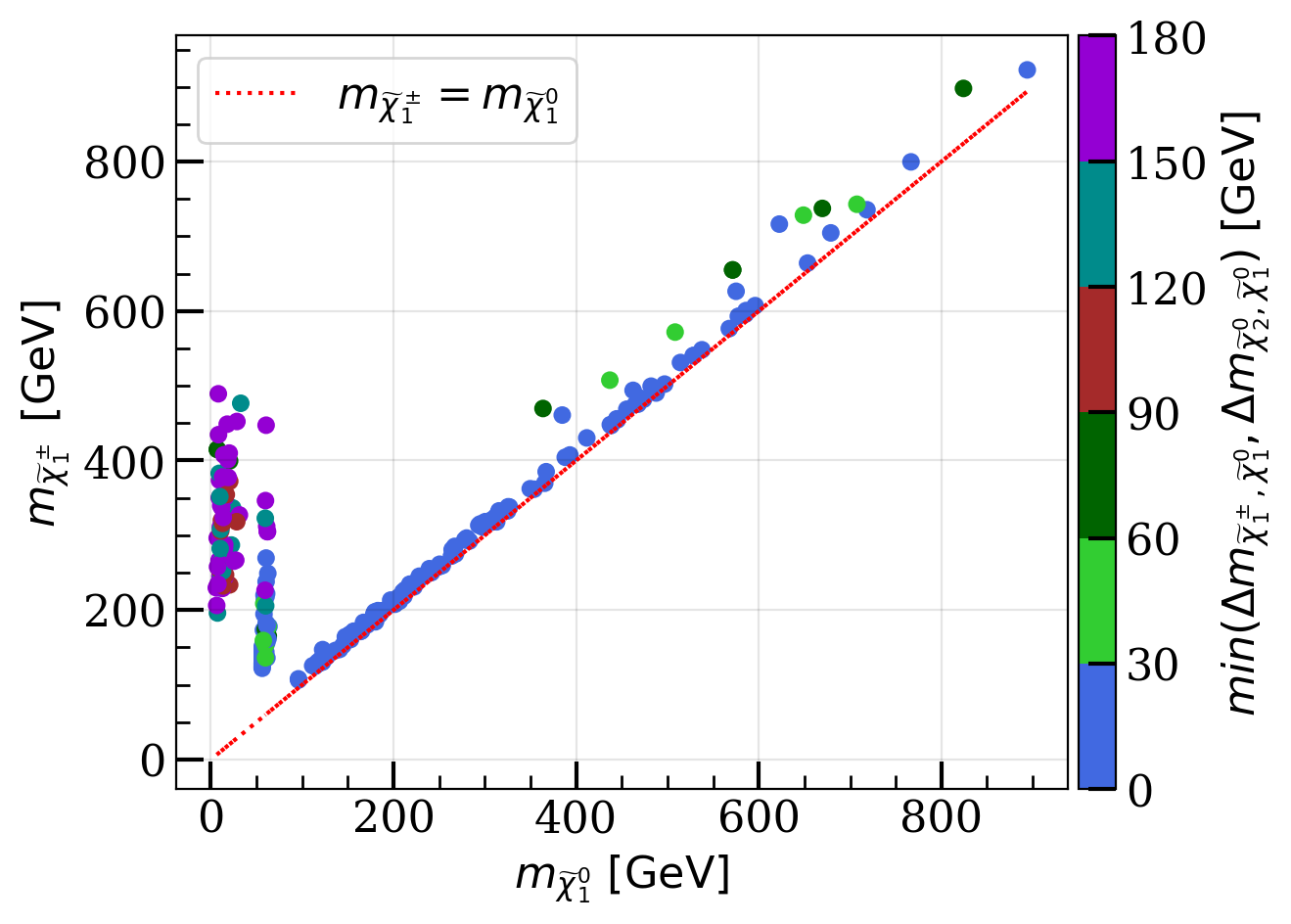}\hfill
    \includegraphics[scale=0.39]{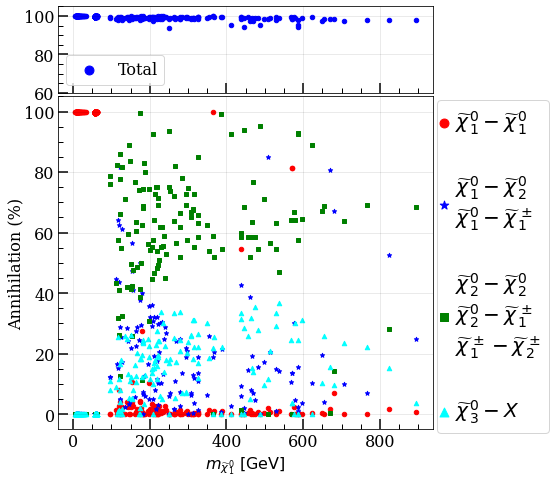}
    \caption{\textit{Left}: Currently allowed parameter points are depicted in the $\mchonepm$ versus $m_{\lspone}$ plane. The $z$-axis represents the minimum of the mass difference between the Singlino-dominated $\lspone$ and the NLSP $\lsptwo/\chonepm$, $\mathrm{min}(\Delta m_{\chonepm,\lspone}, \Delta m_{\lsptwo,\lspone})$. \textit{Right:} The contribution percentage from the different DM annihilation modes in the early universe is shown for the allowed parameter points as a function of $m_{\lspone}$. The points in red represent the contribution from $\lspone-\lspone$ annihilation, including the resonant $s$-channel and $t$-channels. Contributions from co-annihilation with the $\lsptwo$ or $\chonepm$ are shown in blue, while those from NLSP-NLSP-assisted co-annihilation are shown in green. The contribution from co-annihilation with the heavier neutralinos $\lspthree$ is depicted in sky-blue. The top panel represents the sum of contributions from the DM annihilation modes depicted in the lower panel.}
\label{fig:relic_sing}
\end{figure}

As indicated in Eq.~\eqref{eq:gchichiH}, the coupling of the Singlino-dominated neutralinos with the singlet-dominated Higgs bosons relies on $\kappa$, which is typically smaller $\lesssim 0.1$ within our parameter space. Alternatively, it couples with the singlet-dominated Higgs bosons through its Higgsino admixture. At higher LSP masses, $m_{\lspone} \gtrsim 100~$GeV, co-annihilation processes are required to comply with the relic density upper limits. This includes co-annihilation of Singlino-dominated LSP $\lspone$ with the near-degenerate NLSP neutralino $\lsptwo$ or chargino $\chonepm$ containing a non-trivial Higgsino admixture, $\textrm{LSP} + \textrm{NLSP} \to \textrm{SM}$ and the NLSP-NLSP `assisted' co-annihilation, $\textrm{NLSP} + \textrm{NLSP} \to \textrm{SM} $. Typically, when $m_\textrm{LSP} \sim m_\textrm{NLSP}$, the thermal equilibrium between the LSP and the NLSP in the early universe can be maintained by the process $\textrm{LSP} + X \xleftrightharpoons{} \textrm{NLSP} + X^{\prime}$, where $X$ and $X^{\prime}$ are SM fermions. This interaction is mediated via the $t$-channel exchange of gauge bosons or Higgs bosons, with interaction rates governed by $n_\textrm{LSP} n_{X} \langle \sigma v\rangle$, where $n_\textrm{LSP}$ is the number density of the LSP and $\langle \sigma v\rangle$ represents the thermally-averaged cross-section times velocity~\cite{Djouadi:2008uj}. The Higgsino admixture in the LSP $\lspone$, although small but non-negligible, typically boosts these interactions within our parameter space of interest. Likewise, the rate of NLSP-NLSP annihilation process, $\textrm{NLSP} + \textrm{NLSP} \to \textrm{SM}$, scales as $\sim (n_\textrm{NLSP}^{2} \langle \sigma v\rangle)$, where $n_\textrm{NLSP}$ is the number density of the NLSPs. Near the freeze-out temperature, where the number density of SM fermions is several orders of magnitude larger than the NLSP, the interaction rate for the process $\textrm{LSP} + X \to \textrm{NLSP} + X^{\prime}$ is typically larger than $\textrm{NLSP} + \textrm{NLSP} \to \textrm{SM}$. This allows the Singlino-dominated LSP $\lspone$ to annihilate at a rate roughly comparable to that of the Higgsino-dominated NLSP, resulting in the observed under-abundant relic density. In the left panel of Fig.~\ref{fig:relic_sing}, we show the allowed parameter points in the $m_{\lspone}$-$m_{\chonepm}$ plane. The red dashed line represents the mass degeneracy condition required for co-annihilation, $m_{\lspone} = m_{\chonepm}$, and it is observed that the parameter points with $m_{\lspone} \gtrsim 100~$GeV mostly lie along this line. In the $z$-axis, we show the minimum of the mass difference between the $\lspone$ and $\lsptwo/\chonepm$. 

As discussed in Sec.~\ref{sec:constraints}, the chargino mass limits from LEP exclude Winos and Higgsinos up to 103.5~GeV~\cite{OPAL:2003wxm}. The Higgsinos can be further constrained by direct searches at the LHC and direct detection experiments. However, within our parameter space of interest, due to mixing with the gauginos and depending on the relative values of $M_1$ and $M_2$, the lower bounds on Higgsinos are significantly weakened. As a result, we obtain allowed points with Higgsino-dominated charginos and neutralinos as low as $110~$GeV, with $m_{\lspone}$ around the same mass, the criterion required for co-annihilation. It is worth noting that a sub-dominant contribution to the DM annihilation rate arises from $\lspone\lspone$ annihilation into a pair of singlet-like light Higgs bosons: $\lspone \lspone \to H_1 H_1$ when $m_{\lspone} > m_{H_1}$ and $m_{A_1} + m_{H_1} > 2 m_{\lspone}$, $\lspone \lspone \to A_1 A_1$ when $m_{\lspone} > m_{A_1}$ and $m_{A_1} + m_{H_1} > 2 m_{\lspone}$, and $\lspone \lspone \to H_1 A_1$ when $m_{H_1} + m_{A_1} < 2 m_{\lspone}$. These annihilation channels can involve the $s$-channel exchange of the $Z$, $H_1$, or $A_1$ bosons, or the $t$-channel exchange of $\lspone$ and other heavier neutralinos.

We present the relative contribution~(in $\%$) from the different DM annihilation modes for our allowed parameter space points in the right panel of Fig.~\ref{fig:relic_sing}. The relative contribution from $\lspone-\lspone$ annihilation in the $s$- or $t$-channels are shown in red, $\lspone$-$\lsptwo/\chonepm$ co-annihilation is shown in blue, and NLSP-NLSP assisted co-annihilation is shown in green. The contribution from co-annihilation processes involving $\lspthree$ is shown in sky-blue. As discussed previously, resonant $s$-channel $\lspone-\lspone$ annihilation is the dominant mode for DM dilution in the early universe for $m_{\lspone} < m_{H_{SM}}/2$. At higher $\lspone$ masses, co-annihilation with and assisted co-annihilation plays a crucial role in achieving consistency with the relic density limits.


\section{Benchmark Scenario}
\label{sec:future_prospects}

Building on our understanding of the currently allowed parameter space, we now focus on examining the potential of probing these regions at the upcoming HL-LHC. One of the most promising modes to probe the electroweakinos at the LHC is through their direct production, as suggested by a plethora of studies such as \cite{ATLAS:2014ikz, ATLAS:2017qwn, ATLAS:2019wgx, CMS:2021few, CMS:2022sfi, ATLAS:2023act}. However, our approach diverges from the traditional decay topologies, emphasizing NMSSM-specific scenarios aimed at exploring discovery prospects at the HL-LHC. 

For our collider analysis, we examine a scenario from the allowed parameter region where the NLSP and LSP have non-degenerate masses, allowing the NLSP to decay promptly into the LSP. This configuration is observed in the low LSP mass regime, $m_{\lspone} \lesssim m_{Z}/2$, where the primary DM dilution mode is characterized by resonant $s$-channel annihilation through a singlet Higgs boson with roughly twice the mass of the LSP. It is worth noting that in the co-annihilation regime with mass degenerate NLSP and LSP, the NLSPs may become long-lived, as previously explored in \cite{Adhikary:2022pni}. 

\begin{table}[!t]
\centering\scalebox{0.70}{
 \begin{tabular}{|c|l|l|l|l|} 
 \hline
\makecell{BPs}& \makecell{Parameters \\ (GeV wherever applied)}& \makecell{Masses (GeV)} &\makecell{Cross-sections\\ at 14 TeV (fb) } & \makecell{Interesting Processes \&  B.F.s(\%)}\\
\hline\hline
& $\lambda = 0.174$, $\kappa =0.007$,  & $m_{\lspone} = 20.5$, $m_{\lsptwo} = 123.9$      & $\lspthree \chonepm = 194.4$    & $H_1 \to b \bar{b}$ = 82.4, \hfill$H_1 \to \tau^+ \tau^-$ = 10.6,\\
& $A_\lambda = 1384$,$A_\kappa = -70$,   & $m_{\lspthree} = 243$, $m_{\lspfour} = 244$,& $\chonemp \chonepm = 116.4$      & $H_3 \to \lsptwo \lspthree$ = 13.8,\hfill$A_2 \to \lsptwo \lspfour$=12.0,\\ 
& $\tan\beta = 5.15$, $\mu = 229$,    & $m_{\lspfive} = 1028$, $m_{\chonepm} = 233$, & $\lspfour \chonepm = 163.0$       & $\chonepm \to \lspone W$ = 52.3,\hfill$\chonepm \to \lsptwo W$ = 47.7,\\
BP& $A_t = -6348$, $A_b =A_\tau=2000$,  & $m_{\chtwopm} = 1028$, $m_{H_1} = 14.2$,       & $\lspthree \lspfour = 85.04$      & $\chtwopm \to \chonepm Z$ = 25.6,\hfill$\chtwopm \to \lspthree W$ = 24.8,\\
& $A_\mu=0$, $M_1 = 131.4$,             & $m_{H_2} = 125.1$, $m_{H_3} = 1169$,           & $\lsptwo \chonepm = 50.62$       & $\chtwopm \to \chonepm H_2$ =24.1,\hfill$\lsptwo \to \lspone H_1$ = 50.4,\\  
& $M_2= 965.8$, $M_3 = 3663$,           & $m_{H^\pm} = 1170$, $m_{A_1} = 43.8$,         &                                   & $\lsptwo \to \lspone Z = 46.3$,\hfill$\lspthree \to \lspone Z = 31.4$,\\    
&$N_{15}=0.99$, $N_{21}=0.95$,          & $m_{A_2} = 1168$                                &                                   & $\lspthree \to \lsptwo Z =48.9$,\hfill$\lspthree \to \lspone H_2 =16.4$,\\
&$N_{23}=0.26, N_{24}=0.16$,             &                                                   &                                   & $\lspfour \to \lspone Z =45.3$,\hfill$\lspfour \to \lsptwo Z = 14$,\\
&$N_{33}=N_{34}=0.7$,                    &                                                   &                                   & $\lspfour \to \lspone H_2 =20.4$,\hfill$\lspfour \to \lsptwo H_1 =17.2$,\\    
&$N_{41}$=0.3, $N_{43}=N_{44}=0.66$      &                                                   &                                   & $\lspfive \to \lspthree Z =21.4$,\hfill$\lspfive \to \chonepm W =50.4$,\\
&$N_{52}=0.99$, $\Gamma_{A_1} = 4.41 \times 10^{-5}$,                            &                                                   &                                   & $\lspfive \to \lspfour H_2 =19.5$\\   
 & $\Gamma_{H_1} = 9.47 \times 10^{-7}$  & & & \\
\hline
\end{tabular}
}
\caption{The input parameters, neutralino admixtures, masses and branching ratios of the electroweakinos and Higgs bosons, decay length of the light singlet-dominated Higgs bosons, and the cross-section of electroweakino pair production at $\sqrt{s}=14~$TeV, for benchmark point BP.} 
\label{Tab:BP}   
\end{table}
\begin{figure}[!b]
	\centering
	\includegraphics[scale=0.4]{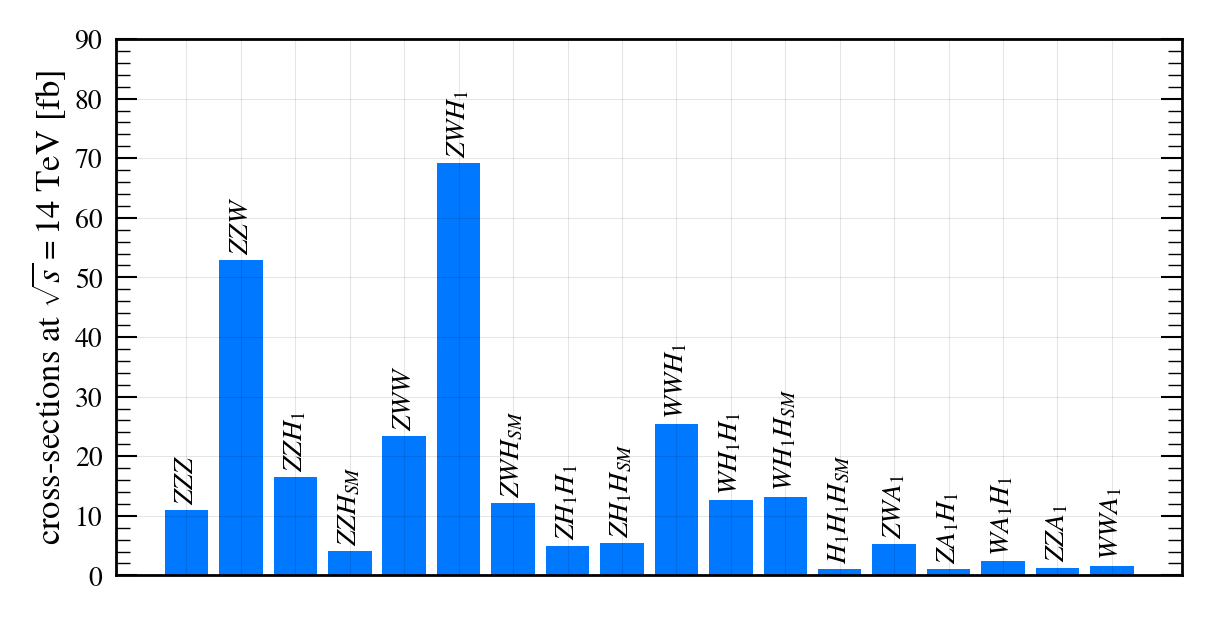}
	\caption{Production rates at the HL-LHC for all possible triple-boson channels from cascade decays of electroweakino pair production corresponding to benchmark point BP~(see Table~\ref{Tab:BP}). Final states with a production rate of $< 1~$fb have been ignored.}
    \label{fig:relative_xs}
\end{figure} 
Within our parameter space of interest, featuring gaugino and Higgsino-dominated heavier neutralinos and charginos, appreciable pair production rates are expected at the LHC. With the Singlino-dominated state serving as the LSP DM, heavier electroweakino states will decay into it in a cascading manner, involving more steps compared to the MSSM scenario. The final state may also involve a light singlet-like Higgs boson, which is a characteristic feature in the NMSSM scenarios. The extended decay chain opens the possibility for probing these scenarios via final states involving triple-boson plus missing energy, $pp \to VVV + \met$~($V = Z, W^{\pm}, H_{SM}, H_1, A_1$), which can be challenging to achieve in the MSSM framework. 

We consider a benchmark point BP from the low LSP mass regime, $m_{\lspone} \lesssim m_{H_{SM}}/2$, in the allowed parameter space. The details of model parameters and the masses of electroweakinos and Higgs bosons are presented in Table~\ref{Tab:BP}. The branching ratios of the heavier charginos and neutralinos are also given, showing $\lsptwo$ with mass $m_{\lsptwo} = 123.9~$GeV as Bino-dominated with significant Higgsino admixture, while $\lspthree, \lspfour$ and $\chonepm$ being Higgsino-dominated with masses 243.2,~244.7 and 233.6~GeV, respectively. This configuration results in $\chonepm \lspthree + \chonepm \lspfour $ having the highest production cross-section at the HL-LHC among all possible chargino-neutralino pair production channels. Alternatively, a benchmark point with a Higgsino-dominated NLSP $\lsptwo$ could potentially lead to higher production rates for $\chonepm \lsptwo$ and larger branching rates into final states with a singlet-like Higgs boson. However, it would preclude the triple-boson~$+\met$ final state. We proceed to analyze the production rates of potential triple-boson~$+\met$ final states arising from $pp \to \chonepm \lspthree + \chonepm \lspfour$ for BP. We display our results in Fig.~\ref{fig:relative_xs}. Interestingly, the triple-boson final state with the highest production cross-section for BP is $ZWH_1$, featuring a light singlet-dominated Higgs boson unique to NMSSM-specific scenarios. We consider this decay channel for our collider analysis in the next section. 


\section{Direct Electroweakino searches in the $ZWH_1 $ channel}
\label{sec:collider}

\begin{figure}[!t]
	\centering
	\includegraphics[scale=0.3]{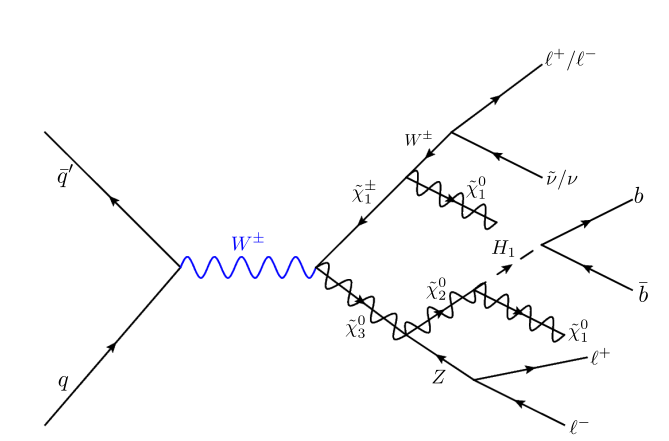}
	\includegraphics[scale=0.3]{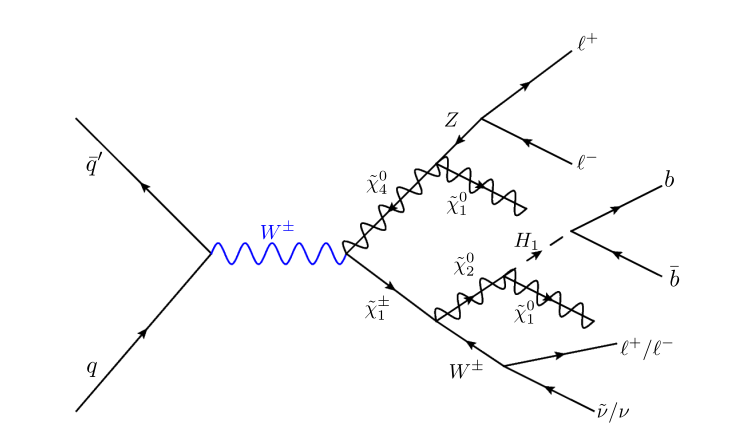}
	\caption{The leading order Feynman diagram for the signal process: $pp \to\chonepm\lspthree + \chonepm \lspfour \to ZW^{\pm}H_1 \to 3 \ell + b\bar{b} + \met$~($\ell = e^{\pm}, \mu^{\pm}, \tau^{\pm}$).}
    \label{fig:feynman_diagram}
\end{figure}

We perform a detailed collider analysis to probe the electroweakinos at the HL-LHC through searches in the $pp \to \chonepm\lspthree + \chonepm \lspfour \to Z W H_1 + \met$ channel, considering the benchmark point BP. As highlighted in Sec.~\ref{eq:cascade_decay}, we choose the $ZWH_1$ channel due to its higher production rate among all potential triple-boson final states arising from the cascade decay of directly produced $\chonepm\lspthree$ or $\chonepm \lspfour$ pairs. The cascade decay chain for the $ZWH_1$ channel is outlined as follows, 
\begin{equation}
\begin{split}
pp &\to \chonepm \lspthree \to (\chonepm \to W^{\pm} \lspone) (\lspthree \to Z (\lsptwo \to H_1 \lspone)) \to Z W^{\pm} H_1 + \met, \\
&\to \chonepm \lspfour \to (\chonepm \to W^{\pm} (\lsptwo \to H_1 \lspone)) (\lspfour \to Z \lspone) \to Z W^{\pm} H_1 + \met.
\end{split}
\label{eq:cascade_decay}
\end{equation}
At benchmark point BP, the tree-level cross-section for the process $pp \to \chonepm\lspthree + \chonepm \lspfour$ is approximately $358~$fb at the $\sqrt{s}=14~$TeV LHC, as detailed in Table~\ref{Tab:BP}. The cascade decay chain in Eq.~\ref{eq:cascade_decay} results in a production rate of approximately $43~$fb for the $ZWH_1$ final state. For the analysis, we consider the dominant decay mode of $H_1$: $H_1 \to b \bar{b}$, with a branching ratio of $Br(H_1 \to b\bar{b}) \sim 82.4\%$. Furthermore, we exclusively focus on the leptonic decay modes of the $Z$ and $W$ bosons due to their cleaner signatures at the detector and to avoid large QCD backgrounds. Thus, the signal process features a final state with three charged leptons, two bottom quarks, along with missing energy owing to the LSP $\lspone$'s: $pp \to \chonepm\lspthree + \chonepm \lspfour \to ZW^{\pm}H_1 \to 3 \ell + b\bar{b} + \met$~($\ell = e^{\pm}, \mu^{\pm}, \tau^{\pm}$). The leading order Feynman diagrams for the signal are presented in Fig.~\ref{fig:feynman_diagram}.  

Dominant contributions to the background arise from top-pair production in association with the gauge bosons, $pp \to t\bar{t}W, t\bar{t}Z$, associated production with the SM Higgs boson, $t\bar{t}H_{SM}$, $ZWb\bar{b}$, $VVV$, $VV+{\textrm jets}$, and $VH_{SM}$ processes, where $V = W^{\pm}, Z$. We consider all possible decay topologies in the background processes that lead to the $3 \ell + b \bar{b} + \met$ final state. Furthermore, sub-dominant contributions from $t H_{SM}$ and $t\bar{t}VV$ processes are also included in the analysis. 

The signal and background events are generated at the parton-level with \texttt{MG5aMC@NLO v2.9.16}~\cite{Alwall:2014hca} utilizing the \texttt{NNPDF23LO} parton distribution function~\cite{Ball:2014uwa} with the \texttt{A14} tune~\cite{ATL-PHYS-PUB-2014-021}. The signal events for BP are generated using the \texttt{NMSSM} model file \cite{Ellwanger:2009dp, Djouadi:2008uw, Skands:2003cj, Allanach:2008qq}, with the masses, couplings, and branching ratios computed using \texttt{NMSSMTools}. Both signal and background events have been generated at the leading order, and higher-order effects are incorporated through $K$ factors. Details of the generation-level cuts implemented in \texttt{MG5aMC@NLO v2.9.16} and cross-sections for the various processes have been included in Appendix~\ref{sec:appendixB}. Subsequently, showering and hadronization is performed using \texttt{Pythia-8}~\cite{Sjostrand:2006za,Sjostrand:2007gs}, and fast detector response simuatation is performed with \texttt{Delphes-3.5.0}~\cite{deFavereau:2013fsa}, utilizing the default HL-LHC configuration card~\cite{HL-LHC_card:Online}. 

We select events containing exactly three isolated leptons $l$ in the final state with transverse momentum $p_{T,\ell} \geq 15~$GeV and pseudorapidity $|\eta| < 4.0$. For electrons and muons, we impose the isolation criteria,
\begin{equation}
\frac{\sum p_{T}^{\Delta R<0.2}}{p_{T,\ell^{\prime}}} < 0.1 \quad \ell^{\prime}=e,\mu, 
\label{eqn:lepton_isolation}
\end{equation}
where $\sum p_{T}^{\Delta R < 0.2}$ represents the sum of the transverse momentum of all other objects within a cone of radius $\Delta R \leq 0.2$ centred around the candidate lepton $\ell^{\prime} = \{e, \mu\}$ carrying transverse momentum $p_{T,\ell^{\prime}}$. Electrons and muons originating from the leptonically decaying $\tau$'s are also subject to the isolation criteria in Eq.~\eqref{eqn:lepton_isolation}. For hadronically decaying tau leptons, $\tau_{h}$, we apply tau-jet tagging efficiencies derived from CMS using the DeepTau algorithm~\cite{CMS:2022prd}. These efficiencies are valid for $\tau_h$ with $p_{T,\tau_h} > 20~$GeV and within $|\eta| < 2.3$. In other kinematic regions, we adopt the default tau-jet tagging efficiencies provided in the HL-LHC configuration card. We also impose $p_{T,\ell_1} > 32~$GeV and $p_{T,\ell_2} > 20~$GeV, where $\ell_{1}$ and $\ell_{2}$ are the isolated leptons with the highest and $2^\textrm{nd}$ highest transverse momentum, respectively. Furthermore, two of the isolated leptons are required to form a same flavor opposite sign~(SFOS) lepton pair with invariant mass close to the mass of the $Z$ boson, $| m_{Z} - m_{\ell\ell}^{SFOS}| < 15~$GeV. The basic event selection cuts discussed until this point are summarized in Table~\ref{tab:cuts_1}. 
\begin{table}[!htb]
    \centering
    \begin{tabular}{c}\hline
    Basic selection cuts \\ \hline\hline
    $n_{\ell^\prime} = 3$ \\
    $p_{T\ell_1,\ell_2,\ell_3}>32,20,15$ GeV \\
    $|m_{Z}-m_{\ell\ell}| < 15~\mathrm{GeV}$\\ \hline 
    \end{tabular}
    \caption{Summary of basic selection cuts.}
    \label{tab:cuts_1}
\end{table}

As discussed previously, a characteristic feature of our signal process is the presence of the light CP-even Higgs boson $H_1$ in the final state, which predominantly decays into a $b\bar{b}$ pair. For our benchmark point BP, $H_1$ with mass $m_{H_1} = 14.2~$ GeV originates from the decay mode $\lsptwo \to H_1 \lspone$, where $\lsptwo$ and $\lspone$ have masses of 123.9 and 20.5~GeV, respectively. The relatively lower mass of $H_1$ coupled with the considerable mass difference between $\{\lspone + H_1\}$ pair and the parent particle $\lsptwo$, results in boosted $H_1$, leading to collimated $ b \bar{b}$ decay products which are typically challenging to resolve. The angular separation between the decay product of the light Higgs, $\Delta R_{b \bar{b}}$, depends on the Higgs mass $m_{H_1}$ and its transverse momentum $p_{T}^{H_1}$ as approximated by $\Delta R_{b\bar{b}} \simeq m_{H_1}/(p_{T}^{H_1}\sqrt{z(1-z)})$ \cite{Butterworth:2008iy}, where $z$ and $1-z$ denote the fractions of momentum carried by the two decay products. Parton-level analysis shows a peak at $\Delta R_{b\bar{b}} \sim 0.2$, which reflects $p_{T,H_1} >> m_{H_1}$. Given the relatively smaller $\Delta R_{b\bar{b}}$, we employ jet substructure techniques to identify the Higgs jet as a fatjet capable of capturing the inherent $b\bar{b}$ pair. We utilize the particle-flow~(PF) algorithm~\cite{ATLAS:2017ghe, CMS:2017yfk} to reconstruct fat-jets using e-flow objects in \texttt{Delphes}~\cite{deFavereau:2013fsa}. These e-flow objects are comprised of PF tracks, including charged particle tracks, and PF towers, including neutral particles and charged particles without any associated tracks, with corrected calorimeter smearings. The electrons and muons are separately included in the PF objects. These objects are given as input to the Cambridge-Achen algorithm~\cite{Dokshitzer:1997in} with the jet radius parameter of $R=0.7$ and the minimum $p_{T}$ threshold of $> 40~$GeV, implemented in the \texttt{FastJet v3.2.1}~\cite{Cacciari:2011ma} setup. The fatjets within $|\eta| < 4.0$ are further groomed using the SoftDrop technique~\cite{Larkoski:2014wba} with free parameters $\beta = 0$ and $z_{cut} = 0.1$, following standard CMS strategies~\cite{CMS:2020poo}. To identify the soft-dropped fatjets as the light Higgs jets, we employ the Mass-drop tagger~\cite{Butterworth:2008iy, Dasgupta:2013ihk} with parameters $\mu = 0.667$ and $y_{cut} > 0.01$, which identifies the two subjets $j_a$ and $j_b$ within the fatjets. $j_a$ and $j_b$ are then matched with the $B$-hadrons~(the truth-level $b$-quarks just before hadronization) with $p_{T} > 10~$GeV and $|\eta| < 2.5$ at the generator-level, using a matching cone of radius $R < min(0.3, \Delta R_{j_a,j_b}/2)$, where $\Delta R_{j_a,j_b}$ represents the separation between $j_a$ and $j_b$. This choice is motivated from Ref.~\cite{Butterworth:2008iy}. If both subjets satisfy these criteria, the fatjet is identified as a tagged light Higgs jet $H^j$. In events with more than one $H^j$, the one with invariant mass closest to $m_{H_1}$ is associated with the light singlet-dominated Higgs boson $H_1$. In this analysis, we impose the criteria, $n_{H^j} = 1$, where $n_{H^j}$ is the number of tagged light Higgs fatjets. For our benchmark BP with $m_{H_1} = 14.2~$GeV, the tagging efficiency of this method is roughly $20\%$. In the remainder of this analysis, we will refer to the tagged $H_1$ jet as $H_1^j$, while the two $b$-quark matched subjets within the $H_1$ jet will be referred to as $b_1$ and $b_2$.    
\begin{figure}[!t]
	\centering
	\includegraphics[scale=0.23]{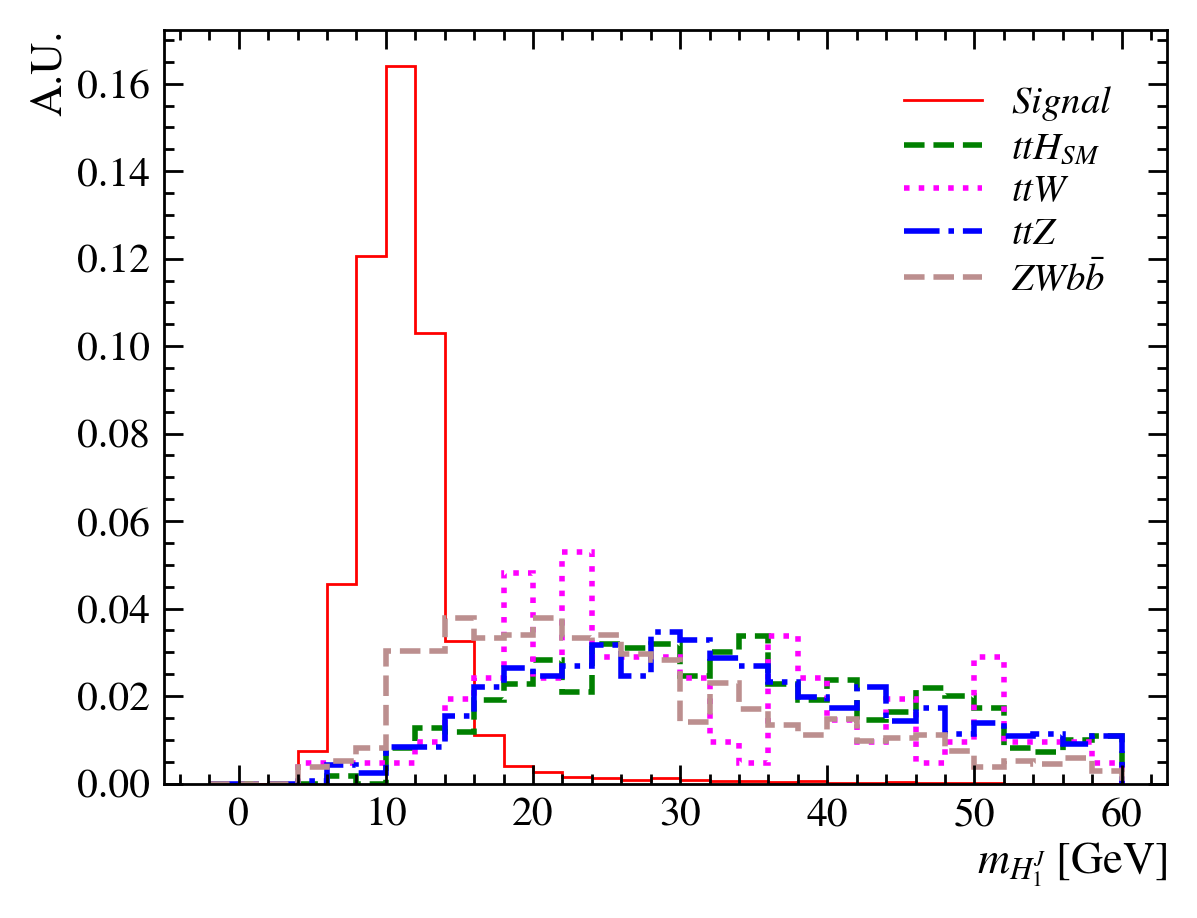}
	\caption{Distribution for the invariant mass of the tagged light Higgs jet $H_{1}^j$ in the $pp \to \chonepm \lspthree + \chonepm \lspfour \to ZWH_1 + \met \to 3\ell + b\bar{b} + \met$ channel at the HL-LHC, corresponding to the benchmark point BP. Distributions for the dominant background processes are also illustrated.}
    \label{fig:mH1}
\end{figure}
In Fig.~\ref{fig:mH1}, we illustrate the mass distribution of the tagged Higgs jet, $m_{H_1^j}$, for the signal process $pp \to \chonepm \lspthree \to ZW^{\pm}H_1 + \met \to 3\ell + b\bar{b} + \met$~(see Eq.~\eqref{eq:cascade_decay}) and the dominant background processes. The $m_{H_1^j}$ distribution for the signal exhibits a sharp peak near $m_{H_1}$, owing to the comparatively smaller decay width of $H_1$~($\Gamma_{H_1} = 9.47 \times 10^{-7}~$GeV), implying good mass resolution. On the other hand, the $m_{H_1^j}$ distributions for the background processes are flatter and do not exhibit any discernible sharp peaks. As such, the $m_{H_1^j}$ observable demonstrates an excellent potential for background reduction. We would like to note that a similar methodology can be extended to tau jets in scenarios where $H_1$ decays to $\tau_h\tau_h$ jets, offering further possibilities to improve the discovery potential of the signal at the LHC.

We perform a cut-based collider analysis by optimizing the selection cuts on 
\begin{itemize}
\item $\Delta R(b_1, b_2)$: $\Delta R$ separation between the two $b$-like subjets associated with $H_{1}^j$,
\item $\met$: missing transverse energy,
\item $\Delta R (\ell_{W}, H_{1_{b_1b_2}})$: $\Delta R$ separation between the non-SFOS lepton $l_{W}$ and $H_{1_{b_1b_2}}$,
\item $\Delta R (Z_{\ell\ell}^{SFOS}, H_{1_{b_1b_2}})$: $\Delta R$ separation between the reconstructed $Z$ boson and $H_{1_{b_1b_2}}$,
\item $H_{T}$: scalar sum of the transverse momentum of the three isolated leptons and the reconstructed $H_{I}^j$ , 
\item $m_{H_1^j}$: invariant mass of the reconstructed $H_1$ fatjet, and
\item $M_{T}(H_{1}^j,\met)$: transverse mass of the Higgs and missing energy system, defined as:
\begin{equation}
M_{T}(H_1^j,\met) = \sqrt{m_{H_1^j}^2 + 2(E_T^{H_1^j}\met - \overrightarrow{\rm p}_{T}^{H_1^b} .\PMETV)}.
\end{equation}
\end{itemize}
We tailor a signal region SR to maximize the signal significance $\sigma_S$ for the signal process in Eq.~\eqref{eq:cascade_decay} for the benchmark point BP~(see Table~\ref{Tab:BP}). The signal significance $\sigma_S$ is defined as, 
\begin{equation}
\sigma_{S} = \frac{S}{\sqrt{B + (B * \delta_{S})^{2}}},
\label{eqn:signal_significance}
\end{equation}
where $S$ and $B$ are the signal and background yields, respectively, and $\delta_{S}$ represents the systematic uncertainty.

\begin{figure}[!htb]
\centering
\includegraphics[width=0.45\textwidth]{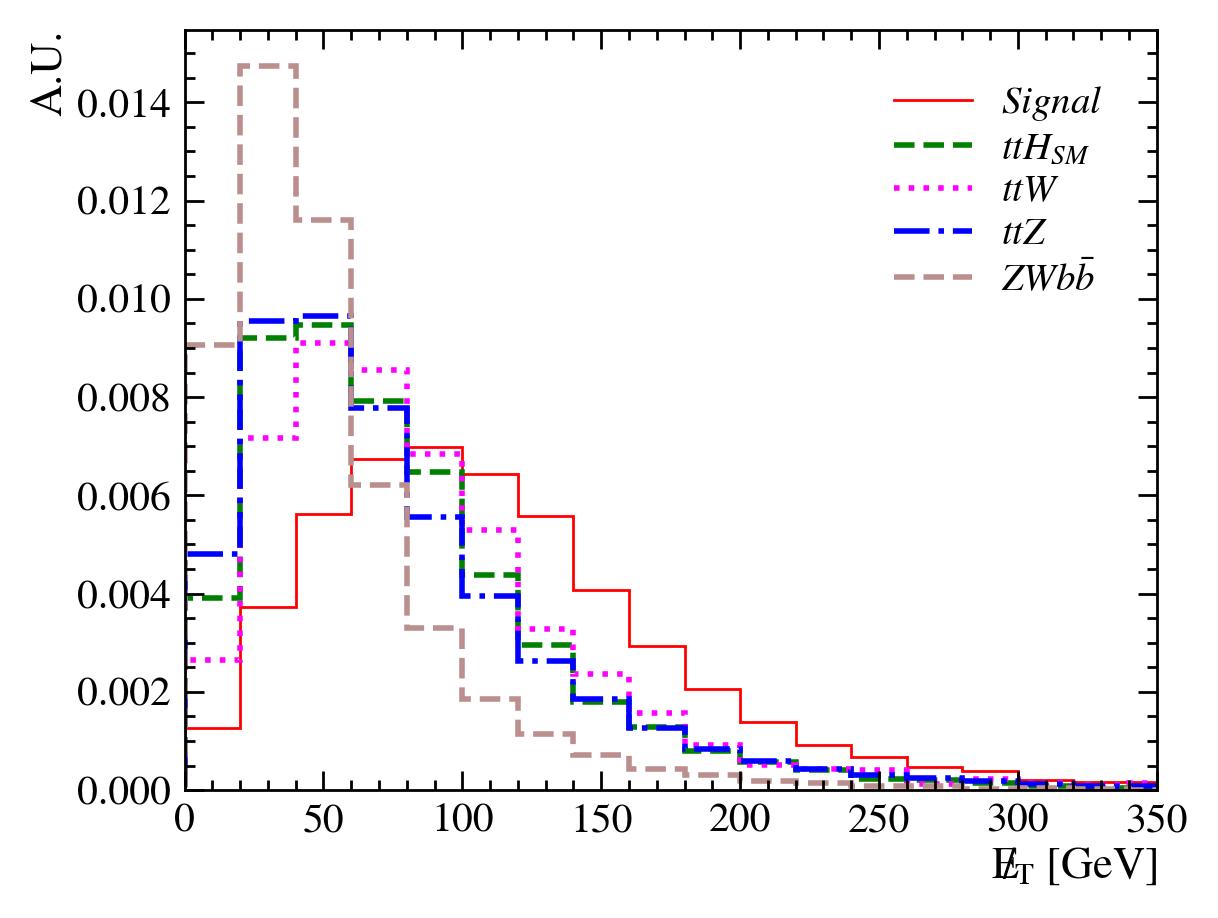}\hspace{0.6cm}
\includegraphics[width=0.45\textwidth]{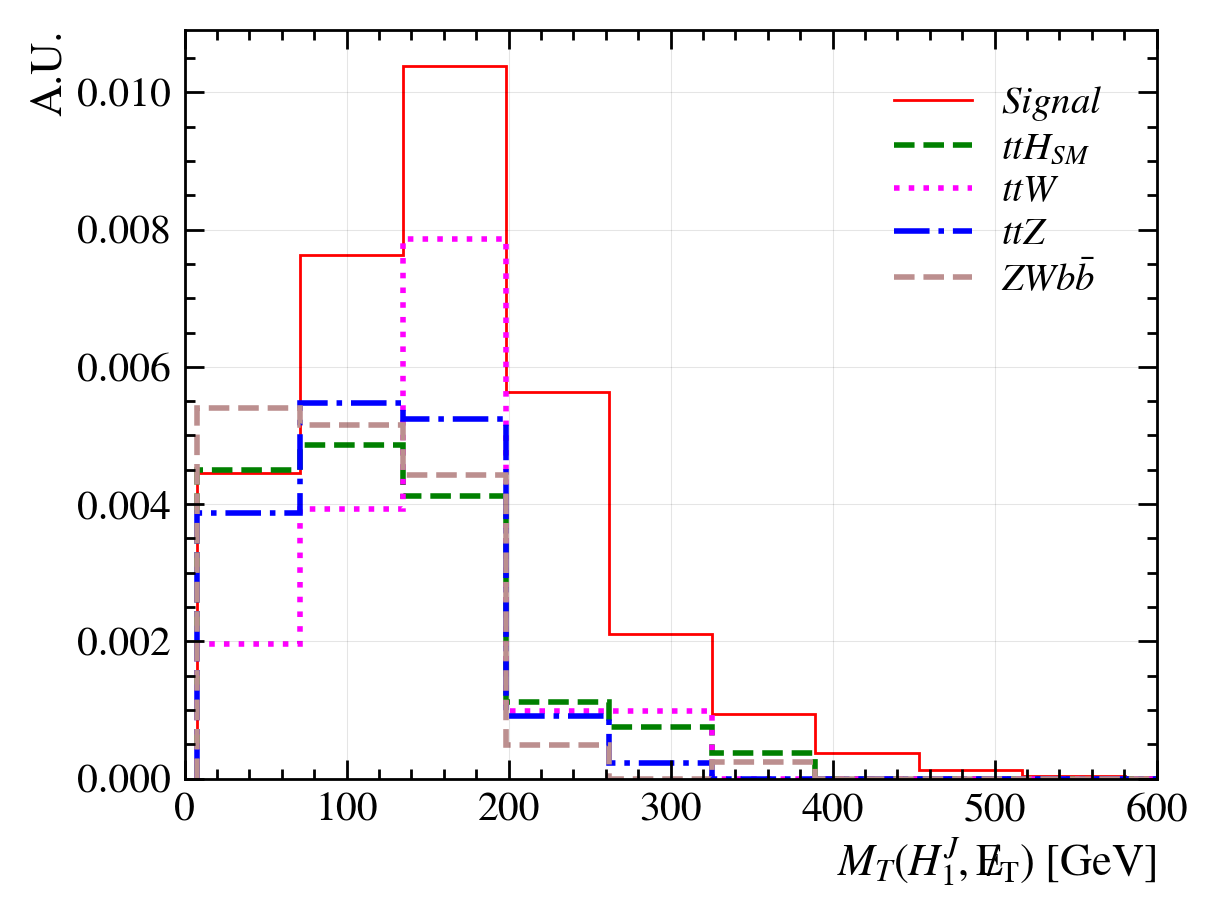}

\caption{Distributions for the missing transverse energy $\met$~(\textit{left}) and the transverse mass of the light Higgs fatjet and missing energy system $M_{T}(H_1^j,\met)$~(\textit{right}) in the  $pp \to \chonepm \lspthree + \chonepm \lspfour \to ZWH_1 + \met \to 3\ell + b\bar{b} + \met$ channel at the HL-LHC, for benchmark point BP. Distributions for the dominant backgrounds are also shown.}
\label{fig:SRA}
\end{figure}

We consider various combinations of selection cuts on the kinematic observables, aiming to maximize $\sigma_S$ for the signal process. It is observed that the subset of observables $\{\met, m_{H_1^j},M_{T}(H_1^j,\met)\}$ leads to the most optimal $\sigma_S$. In Fig.~\ref{fig:SRA}, we illustrate the $\met$ and $M_{T}(H_1^j,\met)$ distributions for the signal process and the dominant backgrounds: $t\bar{t}Z$, $t\bar{t}W$, $t\bar{t}H_{SM}$ and $ZWb\bar{b}$, after imposing the basic selection cuts in Table~\ref{tab:cuts_1}. The $\met$ distribution for the signal falls slowly compared to the backgrounds, with the tails extending to comparatively higher values. The $M_{T}(H_1^j,\met)$ distribution for the signal displays a peak at $M_{T}(H_1^j,\met) \sim 150~$GeV, while the background processes peak at smaller values. The optimized selection cuts, applied sequentially, are listed in Table~\ref{Tab:cut_flow_SA}, along with the signal and background yields at the $\sqrt{s}=14~$TeV LHC with $\mathcal{L}=3000~{fb}^{-1}$. The event yields have been computed as $\sigma \times \textrm{BR} \times \epsilon \times \mathcal{L}$, where $\sigma \times \textrm{BR}$ are the production rates and $\epsilon$ is the efficiency defined as the ratio of the events that pass the cuts to the total number of events. The signal significance values, computed using Eq.~\eqref{eqn:signal_significance}, are also presented for hypothetical scenarios with null and $5\%$ systematic uncertainties. We obtain a signal significance of $\lesssim 1$ upon the imposition of the basic selection cuts. At this stage, the signal over background ratio stands at roughly $10^{-4}$. Imposing $n_{H^j} \geq 1$ and requiring the invariant mass of the $H_1$ fatjet to be less than $18~$GeV resulted in $\sigma_S = 7.1$ and 6.7 for $\delta_S = 0$ and $5\%$, respectively, with $S/B \sim 1$. We further impose a minimum threshold on $M_{T}(H_1^J,\met) > 150~$GeV, which improved $\sigma_S$ to $\sim 7.1$ and $7.0$, respectively. Our results indicate that the benchmark point BP can be probed at the HL-LHC through searches in the $pp \to ZW^{\pm}H_1 \to 3\ell + b\bar{b} + \met$ channel with discovery potential~($\sigma_S > 5\sigma$). 

\begin{table}[!htb]
\centering\scalebox{0.65}{
 \begin{tabular}{|c|c|c|c|c|c|c|c|c|c|c|c|c|} 
 \hline
\multirow{2}{*}{} & \multirow{2}{*}{$SIG$} & \multirow{2}{*}{$ttW$} & \multirow{2}{*}{$ttZ$} & \multirow{2}{*}{$ t t H_{SM}$} & \multirow{2}{*}{$Z W b \bar{b}$} & \multirow{2}{*}{$VVV$} & \multirow{2}{*}{$ V V + jets$} & \multirow{2}{*}{$ V H_{SM}$} &  \multirow{2}{*}{$ t t V V$} & \multirow{2}{*}{$t H_{SM}$}  & \multicolumn{2}{|c|}{$\sigma_s$} \\
\cline{12-13}
& & & & & & & & & & & $\delta_S = 0$ & $\delta_S =5\%$ \\ 
\hline\hline
$\sigma \times \text{BR} \times \mathcal{\ell}$ & 2168 & 1207357 & 2009863  & 1173873 & 160243 & 903398 & 252096000 & 5368320 & 26438 & 169591 &  $0.1$ & $2 \times 10^{-4}$\\[0.2cm]  
Baseline Selection & 402 & 2205 & 23181 & 5290 & 1136 & 5888 & 529371 & 14173 & 211 & 322 & $0.5$ & 0.01 \\[0.2cm]
$\met \geq 60~\mathrm{GeV}$ & 317 & 1372 & 12082 & 2904 & 332 & 2521 & 121223 & 3029 & 138 & 93 & 0.8 & 0.04\\[0.2cm] 
$n_{H^j} = 1$ & 53 & 37 & 285 & 147 & 38 & 14 & 135 & 18 & 4 & $<1$ & 2.0 & 1.2\\[0.2cm] 
$m_{H_j^1} \leq 18$ GeV & 51 & 4 & 24 & 11  & 8 & 0 & 0 & 0 & 0 & 0 & 7.1 & 6.7\\[0.2cm] 
$M_{T} (H_1^j,\met) \geq 150~\mathrm{GeV}$ & 26 & 2 & 7 & 3 & 1 & 0 & 0 & 0 & 0 & 0 & 7.1 & 7.0\\
\hline
\end{tabular}
}
\caption{Signal and background yields in the $pp \to \chonepm \lspthree + \chonepm \lspfour \to ZWH_1 + \met \to 3\ell + b\bar{b} + \met$ channel at the HL-LHC. The yields are presented at each step of the cut-based analysis for the benchmark point BP and the background processes including $ttW, ttZ, tt H_{SM}, ZWb \bar{b}, VVV, VV + jets, V H_{SM}, ttVV$ and $t H_{SM}$. Signal significance at the HL-LHC is also presented for two scenarios: null and $5\%$ systematic uncertainties.} 
\label{Tab:cut_flow_SA}
\end{table}

Through this benchmark analysis, our goal is to emphasize the importance of conducting targeted searches specifically designed for the allowed parameter points. It is also worth noting that the search strategy employed in this analysis could impact other regions of the currently allowed parameter space, especially in the low $m_{\lspone}$ regime where resonant $s$-channel annihilation plays a primary role in achieving the correct or underabundant DM relic density. Investigating the triple-boson final states involving $H_1$ or $A_1$ could be promising in these scenarios, and distinct signal regions could be designed for the parameter points featuring a light Higgs boson by adapting the cuts on the kinematic variables considered in these analyses. We intend to explore these aspects in future work.

\section{Conclusion}
\label{sec:conclusion}

In this work, we focused on exploring the parameter space within the NMSSM framework where a Singlino-dominated neutralino is a viable candidate for thermal dark matter. We concentrated on regions where the neutralino LSP's relic abundance is smaller than the observed cold dark matter relic density. The parameter space of our interest is constrained by various colliders and astrophysical probes, including LEP, rare $B$-meson decays, Higgs signal strengths, BSM Higgs searches, electroweakinos, sparticle searches at the LHC, and DM direct detection experiments. It is observed that a notable fraction of the currently allowed points are well within the reach of future spin-independent direct detection experiments. However, several currently allowed parameter points also lie below the neutrino scattering floor, rendering them inaccessible to future spin-independent direct detection probes. 

In the low mass regime of the LSP neutralino, $m_{\lspone} < m_{Z}/2$, resonant annihilation via the $s$-channel exchange of a light $A_1$ or $H_1$ with mass $\sim 2 m_{\lspone}$ is primarily responsible in achieving consistency with the upper limit on relic density. However, at higher LSP masses, $m_{\lspone} \gtrsim m_{H_{SM}}/2$, the absence of an additional BSM Higgs boson at twice the LSP mass, as prohibited by the Higgs mass rule, restricts efficient DM dilution in the early universe through resonant annihilation via Higgs exchange. In this mass regime, consistency with the relic density constraints is achieved through co-annihilation with the NLSPs such as $\lsptwo$ and $\chonepm$. Assisted co-annihilation, where the $\lspone$ and the NLSP are in thermal equilibrium, also contributes significantly to DM annihilation before freeze-out. 

Having identified the currently allowed parameter space, we then turned our attention to exploring non-conventional search channels to probe them at the HL-LHC. We particularly focus on identifying channels that would be rather challenging to access within the widely explored phenomenological-MSSM scenario. In this context, we explored the prospects of triple boson final states originating from direct electroweakino pair production in Sec.~\ref{sec:future_prospects}. We considered a benchmark point BP from the low LSP mass regime with $m_{\lspone} = 20.5~$GeV from our allowed parameter space, featuring a light singlet-dominated pseudoscalar Higgs boson with twice the LSP mass and scalar Higgs boson at $m_{H_1} = 14.2~$GeV. We analyzed the production rates for all potential electroweakino pair-produced final states involving triple bosons. Our investigation revealed that the $ZWH_1 + \met$ final state achieved the highest production rate for BP. This channel is particularly intriguing due to the presence of the light $H_{1}$, making it specific to the NMSSM framework. We performed a detailed collider analysis in the $pp \to \chonepm \lspthree + \chonepm \lspfour \to ZW(H_1 \to b\bar{b}) + \met \to 3 \ell + b\bar{b} + \met$ channel for BP in Sec.~\ref{sec:collider}. Considering the boosted nature of $H_1$, we perform a fatjet analysis and identify the two $b$-like subjets within the fatjets to effectively identify the Higgs jet. We have presented the details of the optimized selection cuts in Table~\ref{Tab:cut_flow_SA} and obtained a signal significance of $\sim 7 \sigma$, accounting for a systematic uncertainty of $5\%$. This illustrates the potential discovery capability of BP at the HL-LHC through searches in this channel. 

Our results indicate the promising potential of performing benchmark-specific dedicated searches at the colliders to complement the conventional search strategies. It may be worth extending the search channel considered in this work to explore other allowed regions within our parameter space of interest. Furthermore, resonant heavy Higgs production decaying into the same electroweakino pairs could offer additional probes to further boost the sensitivity at the future colliders. We defer these explorations to a future work.

\section*{Acknowledgements}

We would like to thank Aoife Bharucha for helpful discussions during the course of this work. A.A. received support from the French government under the France 2030 investment plan, as part of the Initiative d'Excellence d'Aix-Marseille Université - A*MIDEX. R.K.B.'s work is supported by World Premier International Research Center Initiative (WPI), MEXT, Japan.

\newpage
\appendix
\section{Indirect detection}
\label{sec:appendixA}

\begin{figure*}[!htb]
	\centering
	\includegraphics[scale=0.44]{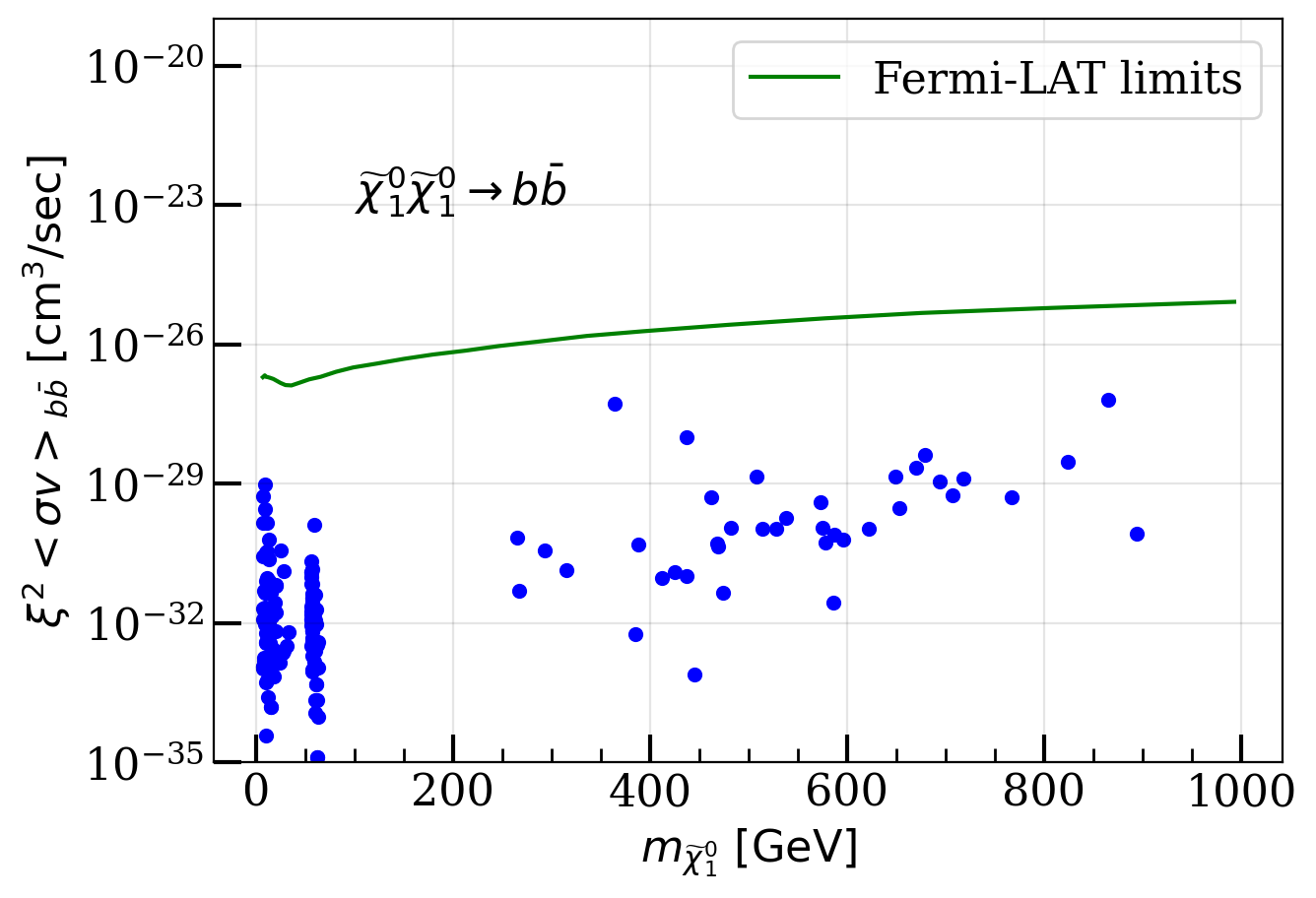}\hfill
	\includegraphics[scale=0.44]{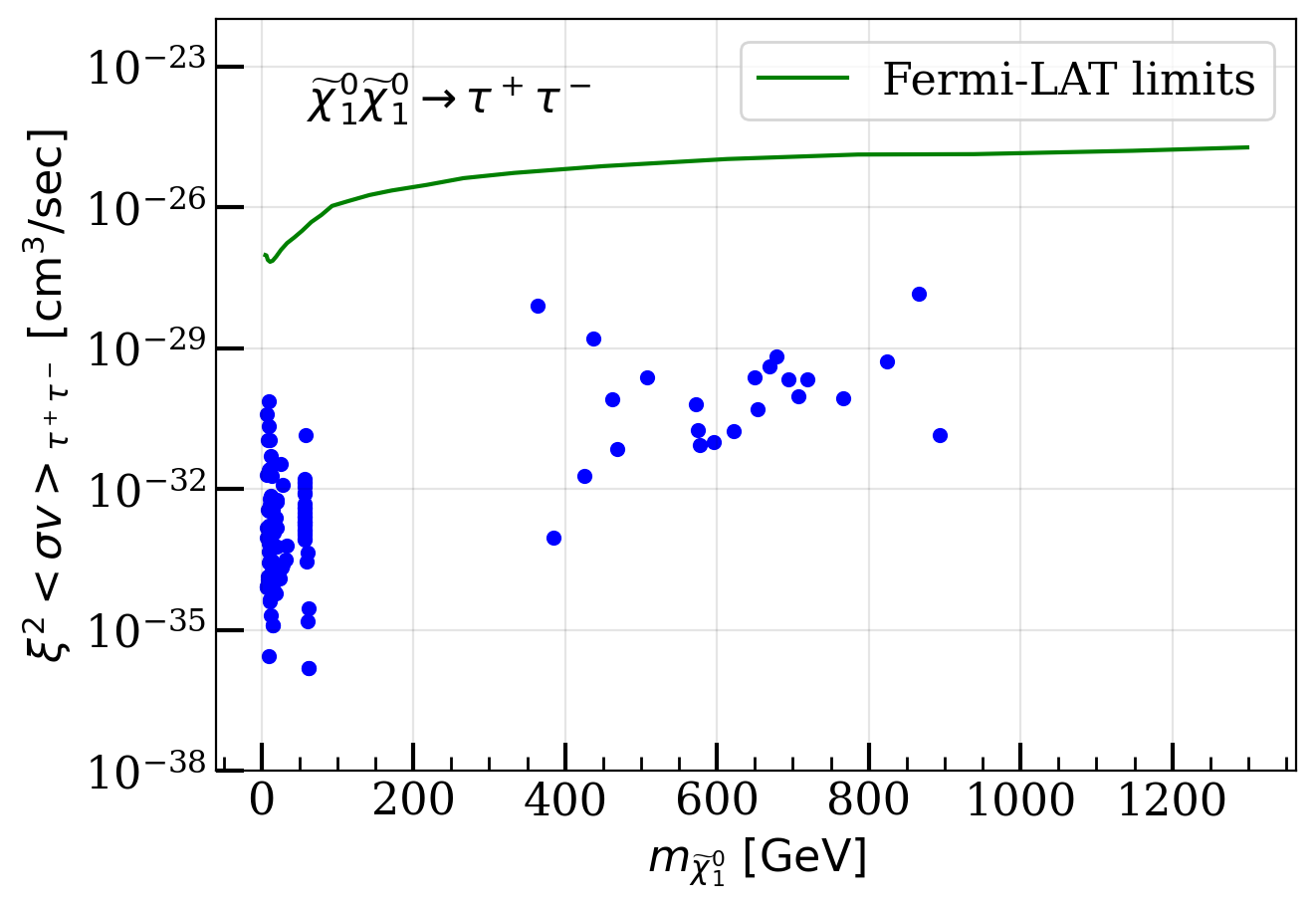}
	\caption{Allowed parameter space points are shown in the plane of $\xi^2$ scaled thermally averaged LSP DM annihilation cross section times velocity in the $b\bar{b}$~(\textit{left}) and $\tau\tau$~(\textit{right}) channels versus the LSP neutralino mass $m_{\lspone}$.}
	\label{fig:ind_det}
\end{figure*}

\section{Summarising the cross sections and generator level cuts for the SM backgrounds}
\label{sec:appendixB}

\begin{table}[h!]
\centering
\scalebox{0.8}{%
\begin{tabular}{|c|c|c|c|c|}

\hline
Backgrounds & Process & \makecell{Generation-level cuts ($\ell=e^\pm,\mu^\pm,\tau^\pm$)\\ (NA : Not Applied)} & Cross section (fb)  \\ \hline\hline
$ttZ$ & $p p \to t \bar{t} Z$ &  NA & 1046.9 \\ \hline
$ttW$ & $p p \to t \bar{t} W^\pm$ &  NA & 628.8\\  \hline
$ttH_{SM}$ & $p p \to t \bar{t} H_{SM}$ &  NA & 611.4\\  \hline
$ZWb \bar{b}$ & $p p \to Z W^\pm b \bar{b}$ & $p_{T, b} > 15 ~\mathrm{GeV}, |\eta_{b}| < 4 , \Delta R_{b,b} > 0.2$ & 83.46 \\  \hline
$VVV$ & $p p \to V V V, (V= W^\pm, Z)$ & NA & 470.5\\  \hline
$VH_{SM}$ & $p p \to V H_{SM}$ & NA & 2796\\  \hline
$ttVV$ & $p p \to t \bar{t} V V$ & NA & 13.77\\  \hline
\multirow{2}{*}{$tH_{SM}$} & \makecell{ model loop SM \\ $p p > W^\pm \to H_{SM} ~~ t/\bar{t} ~~ b/\bar{b} ~\textrm{[QCD]}$} & \multirow{2}{*}{$m_{\ell^+ \ell^-} >30 ~\mathrm{GeV}$} & 3.18\\ \cline{2-2} \cline{4-4} 
 &   \makecell{ model loop SM no~b~mass \\ $p p \to H_{SM} ~~ t/\bar{t} ~~ j ~~\$\$ ~ W^\pm ~\textrm{[QCD]}$} &  & 85.15\\  \hline
$VV + jets$ & \makecell{ $p p \to V V j$ \\ $p p \to V V j j$} & $p_{t, j} > 20 ~\mathrm{GeV}, |\eta_{j}| < 4, \Delta R_{j.j} > 0.2$& 131300 \\

\hline\hline


      \end{tabular}}
           \caption{Generation level cuts and cross-sections for the various Standard Model backgrounds used in the analyses. Cross-sections are obtained from \texttt{Madgraph} LO value multiplied by the $K$ factors.}

\label{app1:1}
          \end{table}

\begingroup
\bibliographystyle{unsrturl}
\bibliography{ref}
\endgroup
\end{document}